\documentclass[prb,onecolumn,superscriptaddress,showpacs]{revtex4}
\usepackage{psfrag,graphicx,stmaryrd,amsmath,amssymb}
\begin{document}

\title{Unconventional charge density wave driven by electron-phonon coupling}

\author{Andr\'as V\'anyolos}
\email{vanyolos@kapica.phy.bme.hu}
\affiliation{Department of Physics, Budapest University of Technology and Economics, 1521 Budapest, Hungary}
\author{Bal\'azs D\'ora}
\affiliation{Department of Physics, Budapest University of Technology and Economics, 1521 Budapest, Hungary}
\author{Attila Virosztek}
\affiliation{Department of Physics, Budapest University of Technology and Economics, 1521 Budapest, Hungary}
\affiliation{Research Institute for Solid State Physics and Optics, PO Box 49, 1525 Budapest, Hungary.}

\begin{abstract}
We report our study on unconventional charge density waves (UCDW) (i.e. a charge density wave with wavevector dependent gap) in pure quasi-one dimensional conductors. We develop a new possible mechanism of establishment of such a low temperature phase, in which the driving force of the phase transition is the electron-phonon interaction with coupling depending on both the momentum transfer ($\mathbf{q}$) and the momentum of the scattered electron ($\mathbf{k}$). Mean field treatment is applied to obtain the excitation spectrum, correlation functions such as the density correlator and the optical conductivity, and the effective mass of the phase excitation. The fluctuation of the order parameter leads to the sliding of the UCDW as a whole. In the absence of impurities, we calculated the effect of this fluctuation on the optical properties. The inclusion of the collective mode significantly alters the optical conductivity, and leads to an effective mass which is nonmonotonic in temperature as opposed to conventional CDWs.
\end{abstract}

\pacs{71.45.Lr, 73.20.Mf, 78.30.-j}
\maketitle

\section{Introduction}
As a result of intense research during the past few decades, much is known about the properties of density waves (DW) possessing a constant $\Delta$ order parameter.\cite{gruner-book} On the other hand, it is also well known that the momentum dependent order parameter $\Delta(\mathbf{k})$ plays an important role in theories of superconductivity (termed unconventional).\cite{anderson,won-maki} Indeed, many unusual features of the high temperature superconductors are understood in terms of a $\mathbf{k}$-dependent single particle gap of $d$-wave symmetry, and therefore the case of $d$-wave superconductors for high-$T_c$ cuprates is now well established.\cite{harlingen,tsuei} Also, most heavy fermion superconductors and organic superconductors appear to be unconventional or nodal with gapless quasiparticle spectrum.\cite{sigrist}

It follows from the success of these developments naturally, that investigating the density wave sector may as well turn out to be a fruitful enterprise. Momentum dependent order parameter in an electron-hole condensate was first introduced in the context of excitonic insulator.\cite{halperin} Since then considerable attention has been focused on investigating unconventional density waves (UDW) under various circumstances.\cite{gulacsi,balazs-sdw,nayak-solo,benfatto,laughlin} A striking feature of these unconventional condensates is that due to the vanishing momentum average of the gap over the Fermi surface $(\langle\Delta(\mathbf{k})\rangle=0)$, these systems are not characterised by spatially periodic modulation of either the charge or the spin density, though the name "density wave" is widely used.\cite{ozaki} This property makes UDW a very likely candidate for those systems where a clear and robust thermodynamic phase transition is not accompanied by an order parameter detectable by conventional means, the situation often referred to as "hidden order".\cite{laughlin}

Possible materials with UDW ground state include: the quasi-two dimensional transition metal dichalcogenid 2H-TaSe$_2$ for which an $f$-wave UCDW was proposed\cite{castro-neto}; the mysterious low temperature micromagnetic phase of the heavy fermion compound URu$_2$Si$_2$ for which an USDW has been suggested originally,\cite{ikeda,ohashi} then later further corroborated\cite{attila}; and the pseudogap phases of (TaSe$_4$)$_2$I,\cite{nemeth,tase4} and the underdoped high-$T_c$ superconductors for which the $d$-wave DW scenario was proposed\cite{laughlin}.

Perhaps, one of the most likely candidate for possessing some kind of CDW of unconventional type in the low temperature phase is the organic conductor $\alpha$-(BEDT-TTF)$_2$KHg(SCN)$_4$. This salt has been investigated experimentally extensively in the past few years,\cite{andres,mori,kartsovnik,basletic,fujita,pouget} and recently, the experimental findings of these latter works have been explained rather well by a quasi-one dimensional UCDW with momentum dependent order parameter, including the threshold electric field,\cite{balazs-threshold,balazs-magneticthreshold,balazs-imperfect} the angular dependent magnetoresistance,\cite{balazs-ucdw-et,balazs-eurlett} and the magnetothermopower and Nernst effect\cite{balazs-ucdw-mtpnernst}.

In this paper, our goal is to develop the theory of a novel type of quasi-one dimensional UCDW, in which the phase transition is governed solely by electron-phonon interaction. A preliminary report of some of our results have already been presented.\cite{ecrys} The article is organised as follows: in section 2 we define our model and develop its mean field theory and thermodynamics. Section 3 is devoted to the optical conductivity and the effective mass of the phase excitation. Our conclusions are given in section 4. 

\section{The model}
\subsection{The coupling and the order parameter}
\noindent Let us start by considering the Hamiltonian of the interacting electron-phonon system of our quasi-one dimensional conductor
\begin{equation}\label{hamiltonian}
H=\sum_{\mathbf{k},\sigma}\epsilon(\mathbf{k})c_{\mathbf{k},\sigma}^+c_{\mathbf{k},\sigma}
+\sum_{\mathbf{q},\lambda}\omega_\lambda(\mathbf{q})a_{\mathbf{q},\lambda}^+a_{\mathbf{q},\lambda}
+\frac{1}{\sqrt{N}}\sum_{\mathbf{k,q},\lambda,\sigma}D_\lambda(\mathbf{k,q})c_{\mathbf{k+q},\sigma}^+c_{\mathbf{k},\sigma}
(a_{\mathbf{q},\lambda}+a_{-\mathbf{q},\lambda}^+),
\end{equation}
where the one particle energy $\epsilon(\mathbf{k})$ for the electrons is given by the usual highly anisotropic $(t_a\gg t_b\gg t_c)$ nearest neighbor tight-binding formula measured from the chemical potential
\begin{equation}\label{kinetic}
\epsilon(\mathbf{k})=-2t_a\cos(ak_x)-2t_b\cos(bk_y)-2t_c\cos(ck_z)-\mu.
\end{equation}
Henceforth, the most conducting crystal axis $x$ will often be called the chain direction. Our system is based on an orthorhombic lattice with lattice constants $a,b,c$, and with one atom per unit cell. The three acoustic phonon branches that the crystal possesses, labelled by the polarization index $\lambda$, have all been taken into account in Eq.~\eqref{hamiltonian}. As usual, $c_{\mathbf{k},\sigma}$ is the annihilation operator of a Bloch electron with momentum $\mathbf{k}$ and spin $\sigma$ in a single band, while $a_{\mathbf{q},\lambda}$ is the same for a free phonon with energy $\omega_\lambda(\mathbf{q})$, momentum $\mathbf{q}$ and polarization $\lambda$. The interaction matrix element in Eq.~\eqref{hamiltonian} is given by
\begin{equation}\label{coupling}
\frac{1}{N}D_\lambda(\mathbf{k,q})=\frac{1}{\sqrt{2M\omega_\lambda(\mathbf{q})}}
\int\text{d}^3r\,\psi_\mathbf{k+q}^*(\mathbf{r})\mathbf{e}_\lambda(\mathbf{q})
[-\nabla U_\text{at}(\mathbf{r})]\psi_\mathbf{k}(\mathbf{r}),
\end{equation}
with $U_\text{at}$ being the atomic potential, $N$ is the number of unit cells in the sample, and $\mathbf{e}_\lambda(\mathbf{q})$ denotes the unit vector of polarization $\lambda$ and momentum $\mathbf{q}$. The $\mathbf{k}$-dependence of the coupling - emanating from second quantization on Bloch-basis $(\psi_\mathbf{k}(\mathbf{r}))$ - turns out to be crucial in order to form an unconventional charge density wave (UCDW). Indeed, it is clear from Eq.~\eqref{coupling}, that choosing plane waves $(\sim e^{i\mathbf{kr}})$ for the one electron basis, one ends up with the well known Fr\"ochlich model\cite{frohlich,lra} of the conventional charge density waves (CDW) with coupling simplified to the Fourier component (at wavenumber $\mathbf{q}$) of the atomic potential. This approximation would therefore lack any kind of unconventionality, which can only originate from the $\mathbf{k}$-dependence of the electron-phonon coupling.

Let us now proceed with the mean field treatment of the Hamiltonian in Eq.~\eqref{hamiltonian}. We introduce the phononic operator $\phi_\lambda(\mathbf{q})=a_{\mathbf{q},\lambda}+a_{-\mathbf{q},\lambda}^+$. Within mean field theory, the operator is replaced by its expectation value $\langle\phi_\lambda(\mathbf{q})\rangle$, which differs from zero only if the momentum is set to $\mathbf{q}=\pm\mathbf{Q}$, with $\mathbf{Q}=(2k_F,\pi/b,\pi/c)$ being the best nesting vector. With this, the electronic part of Eq.~\eqref{hamiltonian} simplifies to a quadratic Hamiltonian in the electronic operators, and one obtains
\begin{equation}\label{hamiltonian-meanfield}
H_\text{MF}=\sum_{\mathbf{k}}\Psi^+(\mathbf{k})(\xi(\mathbf{k})\rho_3+\Delta'(\mathbf{k})\rho_1
-\Delta''(\mathbf{k})\rho_2)\Psi(\mathbf{k}).
\end{equation}
Here $\rho_i$ stand for the Pauli matrices acting on the space of left-, and right-going electrons, $\Psi^+(\mathbf{k})=(c^+_{\mathbf{k}\shortuparrow},c^+_{\mathbf{k-Q}\shortuparrow},c^+_{\mathbf{k}\shortdownarrow},c^+_{\mathbf{k-Q}\shortdownarrow})$ is the four component spinor operator, $\xi(\mathbf{k})$ denotes the linearized electron spectrum in $k_x$ around $\pm k_F$, the Fermi surface: $\xi(\mathbf{k})=v_F(|k_x|-k_F)-2t_b\cos(bk_y)-2t_c\cos(ck_z)$, and finally $\Delta(\mathbf{k})=\Delta'(\mathbf{k})+i\Delta''(\mathbf{k})$ is the decomposition of the order parameter into real and imaginary parts. The order parameter itself reads as
\begin{equation}\label{order-definition}
\Delta(\mathbf{k})=\frac{1}{\sqrt{N}}\sum_\lambda\langle\phi_\lambda(\mathbf{Q})\rangle D_\lambda^*(\mathbf{k,-Q}).
\end{equation}
Now the diagonalization of Eq.~\eqref{hamiltonian-meanfield} is straigthforward: for the excitation energies we have $E_{\pm}(\mathbf{k})=\pm E(\mathbf{k})$, $E(\mathbf{k})=\sqrt{\xi(\mathbf{k})^2+|\Delta(\mathbf{k})|^2}$, which is the usual two band quasiparticle spectrum known for example from the theory of unconventional spin-, and charge density waves driven by purely electronic correlations.\cite{balazs-sdw} The thermodynamics of the system in the low temperature phase is determined solely by the momentum dependence of $E(\mathbf{k})$, and the possible node structure (line or point nodes) on the Fermi surface, if there are any. Thus in the followings we explore the momentum dependence of the order parameter. Making use of the Bloch\,-\,Wannier transformation for the electronic wave functions, $\psi_\mathbf{k}(\mathbf{r})=1/\sqrt{N}\sum_{\mathbf{R}}e^{i\mathbf{kR}}\varphi(\mathbf{r-R})$, we change representation in Eq.~\eqref{coupling} to Wannier basis and obtain
\begin{align}
D^*_\lambda(\mathbf{k,-Q})&=g_\lambda\sum_\mathbf{R}A_\lambda(\mathbf{R})e^{-i\mathbf{kR}},\label{d-fourier}
\end{align}
where
\begin{equation}
A_\lambda(\mathbf{R})=\mathbf{e}_\lambda(\mathbf{Q})\int\text{d}^3r\,\varphi(\mathbf{r})
\left(\sum_\mathbf{R'}e^{i\mathbf{QR'}}[-\nabla U_\text{at}(\mathbf{r-R'})]\right)
\varphi^*(\mathbf{r-R}),\label{overlap-integral}
\end{equation}
and $g_\lambda=(2M\omega_\lambda(\mathbf{Q}))^{-1/2}$. Equation~\eqref{d-fourier} is clearly a Fourier expansion of the coupling in its argument $\mathbf{k}$ in the Brillouin zone harmonics, with coefficients being overlap integrals of Wannier orbitals localized on different sites. Since our system is a tight-binding solid (see Eq.~\eqref{kinetic}), these are considered to be exponentially small if the spacing is large between the two orbitals. Therefore it is a reasonable approximation to retain only the on-site $(\mathbf{R}=0)$ and the nearest neighbor terms $(\mathbf{R}=\pm\mathbf{a},\pm\mathbf{b},\pm\mathbf{c})$. Indeed, as is easily seen from Eqs.~\eqref{order-definition} and~\eqref{d-fourier}, in order to keep the possibility of an UCDW formation with momentum dependent order parameter, one has to go beyond the on-site term. This situation is similar to the development of a CDW in an interacting electron system: in the simple Hubbard model the interaction $U<0$ is not able to drive the system into an unconventional charge density wave ground state either, and one has to include two center exchange integrals as well to facilitate an unconventional condensate.\cite{ozaki,ikeda,balazs-sdw} With all this
\begin{equation}
D^*_\lambda(\mathbf{k,-Q})=D^\lambda_0+D^\lambda_1\cos(bk_y)+D^\lambda_2\sin(bk_y)+D^\lambda_3\cos(ck_z)+D^\lambda_4\sin(ck_z),\label{coupling-expanded}
\end{equation}
where we have set $k_x=k_F$, since the relevant $\mathbf{k}$-dependence of both the coupling and the gap is confined to a narrow region near the Fermi sheet at $+k_F$. The coefficients are therefore given by
\begin{align}
D^\lambda_0&=g_\lambda[A_\lambda(0)+A_\lambda(\mathbf{a})e^{-ik_Fa}+A_\lambda(-\mathbf{a})e^{ik_Fa}],\\
D^\lambda_1&=g_\lambda[A_\lambda(\mathbf{b})+A_\lambda(-\mathbf{b})],\\
D^\lambda_2&=-ig_\lambda[A_\lambda(\mathbf{b})-A_\lambda(-\mathbf{b})],\\
D^\lambda_3&=g_\lambda[A_\lambda(\mathbf{c})+A_\lambda(-\mathbf{c})],\\
D^\lambda_4&=-ig_\lambda[A_\lambda(\mathbf{c})-A_\lambda(-\mathbf{c})].
\end{align}
Consequently, the gap will be of the form
\begin{equation}\label{order-expanded}
\Delta(\mathbf{k})=\Delta_0+\Delta_1\cos(bk_y)+\Delta_2\sin(bk_y)+\Delta_3\cos(ck_z)+\Delta_4\sin(ck_z),
\end{equation}
with
\begin{equation}\label{order-coeff}
\Delta_i=\frac{1}{\sqrt{N}}\sum_\lambda\langle\phi_\lambda(\mathbf{Q})\rangle D^\lambda_i,\quad\quad i=0,\dots,4.
\end{equation}
The structure of the gap function in Eq.~\eqref{order-expanded} makes it clear, that there are line nodes on the Fermi surface in any of the latter four unconventional cases (e.g.: $\Delta_2\sin(bk_y)=0$, $k_z$ varies freely). The quasiparticles around these nodes, often termed nodal in the terminology of superconductivity, will determine the nature of the thermodynamics. Since the gap function itself is the same as the one in quasi-one dimensional electronic UDWs\cite{balazs-sdw} (either spin or charge), the thermodynamics of a one-component (with only one type of gap amplitude being finite among the four possible unconventional cases in Eq.~\eqref{order-expanded}, e.g.: $\Delta(\mathbf{k})=\Delta_2\sin(bk_y)$) phononic UCDW is therefore identical to those of either an electronic UDW,\cite{balazs-sdw} or a $d$-wave superconductor\cite{maki-dwave} (in spite of the different topology of their Fermi surfaces). 

\subsection{The gap equation}
In this subsection, we shall now proceed with the derivation of the self-consistency condition for the order parameter $\Delta(\mathbf{k})$ known as the gap equation. Let us consider the equation of motion of the phonon operator $\phi_\lambda(\mathbf{Q})$ introduced in the previous subsection
\begin{equation}\label{eom}
-\frac{d^2}{dt^2}\phi_\lambda(\mathbf{Q})=\omega_\lambda(\mathbf{Q})^2\phi_\lambda(\mathbf{Q})+2\omega_\lambda(\mathbf{Q})B_\lambda(-\mathbf{Q}),\quad\quad\lambda=1,2,3
\end{equation}
with
\begin{equation}\label{electron-operator}
B_\lambda(-\mathbf{Q})=\frac{1}{\sqrt{N}}\sum_{\mathbf{k},\sigma}D_\lambda(\mathbf{k,-Q})c_{\mathbf{k-Q},\sigma}^+c_{\mathbf{k},\sigma}.
\end{equation}
Taking the expectation value of Eq.~\eqref{eom} over the full Hilbert space, the anomalous electron-hole expectation value appearing in Eq.~\eqref{electron-operator} can be easily calculated in the low temperature phase, and using Eq.~\eqref{order-definition} we get
\begin{equation}\label{B}
\begin{split}
\langle B_\lambda(-\mathbf{Q})\rangle_\text{MF}&=\sum_{\lambda'}\langle\phi_{\lambda'}\rangle\left(-\frac{1}{N}\sum_{\mathbf{k},\sigma}D_\lambda(\mathbf{k,-Q})D_{\lambda'}^*(\mathbf{k,-Q})\frac{\tanh(\beta E(\mathbf{k})/2)}{2E(\mathbf{k})}\right)\\
&\equiv\sum_{\lambda'}\langle\phi_{\lambda'}\rangle\chi_{\lambda\lambda'}.
\end{split}
\end{equation}
At this point it is important to call the attention to the fact that since the susceptibility matrix $\chi_{\lambda\lambda'}$ in Eq.~\eqref{B} is clearly not diagonal, the three equations for the corresponding phonon modes are thus coupled together in Eq.~\eqref{eom}. This can be easily understood: although there is no direct interaction between the different phonon modes, the electron system due to its coupling to all branches acts as a medium, and mediates an effective phonon-phonon interaction. Applying the ansatz $\langle\phi_\lambda(t)\rangle=\langle\phi_\lambda(0)\rangle e^{-i\Omega t}$ to Eq.~\eqref{eom} averaged over the full Hilbert space, the eigenfrequencies $\Omega_\mu^2$ $(\mu=1,2,3)$ of the new, effectively noninteracting quasiparticles are obtained from the following eigenvalue problem
\begin{equation}\label{diag}
\Omega^2\langle\phi_\lambda\rangle=\sum_{\lambda'}[\omega_\lambda^2\delta_{\lambda\lambda'}+2\omega_\lambda\chi_{\lambda\lambda'}]\langle\phi_{\lambda'}\rangle,\quad\quad\lambda=1,2,3.
\end{equation}
Due to the temperature dependence of $\chi_{\lambda\lambda'}$, the matrix in the square brackets in Eq.~\eqref{diag} may develop a zero eigenvalue signalling a soft mode at sufficiently low $T$. The point where this takes place defines the transition temperature $T_c$ of the charge density wave instability. The gap equation is essentially equivalent to the softening condition of any of the three new phonon modes below the transition temperature $(\Omega(T<T_c)=0)$
\begin{equation}\label{diag2}
0=\sum_{\lambda'}[\omega_\lambda\delta_{\lambda\lambda'}+2\chi_{\lambda\lambda'}]\langle\phi_{\lambda'}\rangle,\quad\quad\lambda=1,2,3.
\end{equation}
Adding up the three equations for $\lambda$ in Eq.~\eqref{diag2}, after some straightforward algebra one obtains the gap equation for $\Delta(\mathbf{k})$ in the familiar form\cite{balazs-sdw,maki-nodal}
\begin{align}
\Delta(\mathbf{k})&=\frac{1}{N}\sum_\mathbf{k'}P(\mathbf{k,k'})\Delta(\mathbf{k'})\frac{\tanh(\beta E(\mathbf{k'})/2)}{2E(\mathbf{k'})},\label{gap-equation}\\
P(\mathbf{k,k'})&=\sum_\lambda\frac{4}{\omega_\lambda(\mathbf{Q})}D_\lambda^*(\mathbf{k,-Q})D_\lambda(\mathbf{k',-Q}).\label{interaction}
\end{align}
Once the gap equation is solved, comparing Eqs.~(\ref{gap-equation}-\ref{interaction}) with Eq.~\eqref{order-definition} yields the soft mode amplitude as
\begin{equation}\label{soft-mode}
\langle\phi_\lambda\rangle=\frac{1}{\sqrt{N}}\frac{4}{\omega_\lambda(\mathbf{Q})}\sum_{\mathbf{k}}D_\lambda(\mathbf{k},-\mathbf{Q})\frac{\Delta(\mathbf{k})}{2E(\mathbf{k})}\tanh\left(\frac{\beta E(\mathbf{k})}{2}\right).
\end{equation}

At this point, it is worth noting, that one can deduce the very same gap equation by minimizing the thermodynamic potential of the system
\begin{equation}
\Omega(\mu,T;\{\langle\phi_\lambda\rangle\})=\frac{1}{2}\sum_\lambda\omega_\lambda(\mathbf{Q})|\langle\phi_\lambda\rangle|^2-\frac{2}{\beta}\sum_{\mathbf{k},\alpha=\pm}\ln(1+e^{-\alpha\beta E(\mathbf{k})})
\end{equation}
with respect to its undetermined variables $\langle\phi_\lambda\rangle$.

Before we go on to the next subsection dealing with the solution of the gap equation, we find it necessary and instructive to summarize what the theory has provided us with so far. Up to this point our calculations have been completely general in the sense, that no assumptions were made on the particular form of the electron-phonon coupling $D_\lambda(\mathbf{k},\mathbf{q})$. The primary formula in Eq.~\eqref{order-definition} shows that the momentum dependence of the gap stems solely from the coupling. If it is unconventional, then the averaged value over the Fermi surface vanishes ($\langle\Delta(\mathbf{k})\rangle=0$) leading to a spatially homogeneous density of the electronic charge. This state of affairs might remind one to the similar result found in unconventional spin-, and charge density waves governed purely by electronic correlations.\cite{balazs-modern} Although there are similarities in many aspects between the two types of condensates, there is one point that is completely different in nature in the phononic UCDWs under focus, and must be emphasized by all means. Namely, though there is no spatial modulation of the electronic density at all, the Peierls distortion of the underlying lattice with periodicity $\mathbf{Q}$ is still present (note the finite soft mode amplitude in the low temperature phase in Eq.~\eqref{soft-mode}), which could be detected by usual means, like x-ray scattering. This feature is to be contrasted with that of an electronic DW (either spin or charge), where the density wave instability is caused purely by the electron-electron interaction, and the lattice is not affected at all. Further, if the system favours an unconventional ground state, then there is no modulation either in spin or charge. Thus these systems are often referred to as "systems with hidden order", and are consequently promising candidates for systems where robust thermodynamic signals of a phase transition are seen without any order parameter. 

On the other hand, the static deformation of the lattice is a charasterictic feature of the conventional charge density waves with constant gap,\cite{gruner-book} and in such systems it is always accompanied by the electronic charge oscillation. Now we face a different and novel  situation, namely: in a phononic UCDW these two phenomena need not go together.

Regarding all these, we conclude that the notion "hidden order" might not be applicable to the phononic UCDW, because as entering the low temperature phase a new periodicity with wavevector $\mathbf{Q}$ develops. Nevertheless, the model exhibits all peculiar properties of the aforementioned UCDW that were applied to explain the experimental findings measured on the $\alpha$-(BEDT-TTF)$_2$KHg(SCN)$_4$ salt.\cite{balazs-threshold,balazs-magneticthreshold,balazs-imperfect,balazs-eurlett,balazs-ucdw-et,balazs-ucdw-mtpnernst} Moreover, the extra feature of the distorted lattice might give an explanation to a recent x-ray study performed on this salt,\cite{pouget} in which enhanced structural modulation is found below $T\sim10$K suggesting a coupling to the electronic degrees of freedom.

In the followings, the next subsection will be devoted to the solution of Eq.~\eqref{gap-equation}.

\subsection{Competing phases in the ground state}
\begin{figure}
\psfrag{x}[t][b]{$T/T_c$}
\psfrag{y}[b][t]{$h(T)/h(0)-h_0(T)/h_0(0)$}
\psfrag{x1}[t][b]{$T/T_c$}
\psfrag{y1}[b][b]{$h_0(T)/h_0(0)$}
\psfrag{c}[Bl][Bl][0.8][0]{$D_0/D_2=0.6$}
\psfrag{e}[Bl][Bl][0.8][0]{$1.3$}
\includegraphics[width=8.2cm,height=6cm]{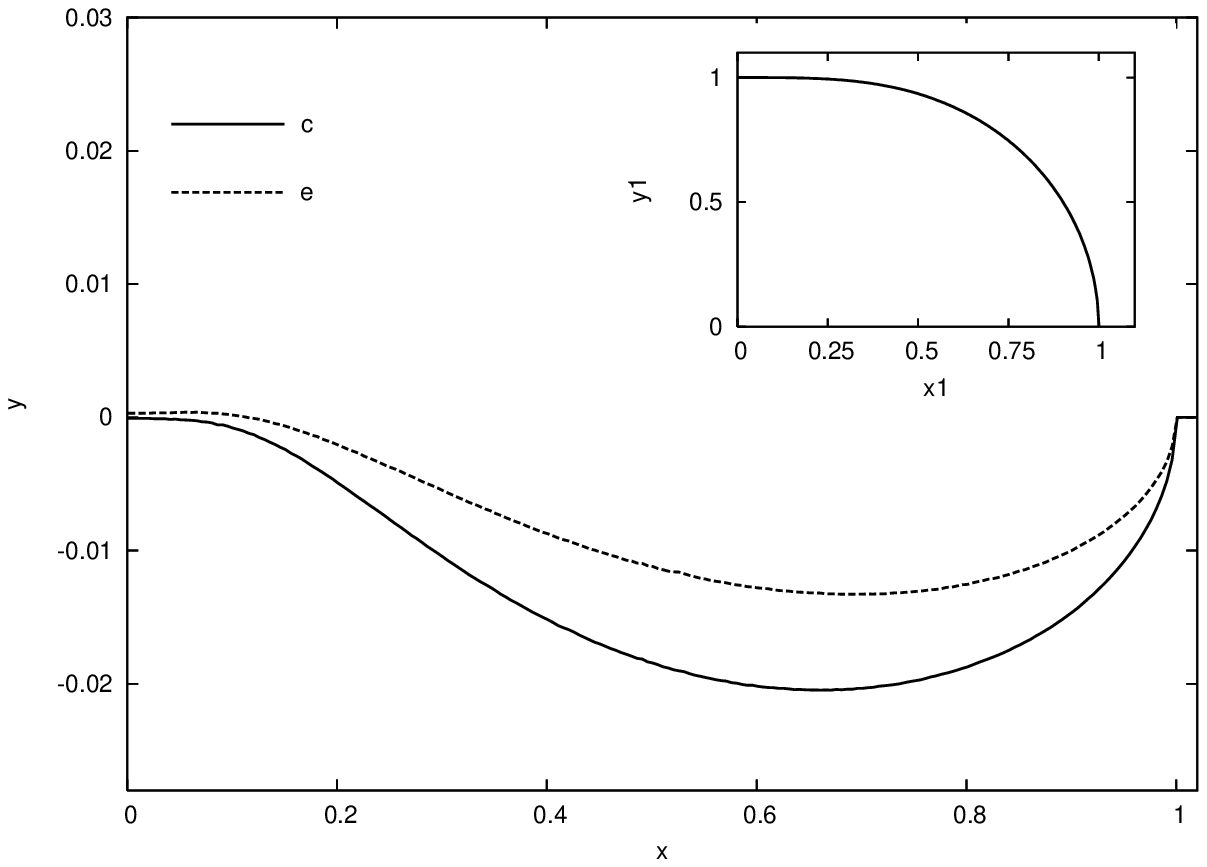}\hfill
\psfrag{x}[t][b]{$|\omega|/\Delta_2$}
\psfrag{y}[b][t]{$\rho(\omega)/\rho_0(0)$}
\includegraphics[width=8.2cm,height=6cm]{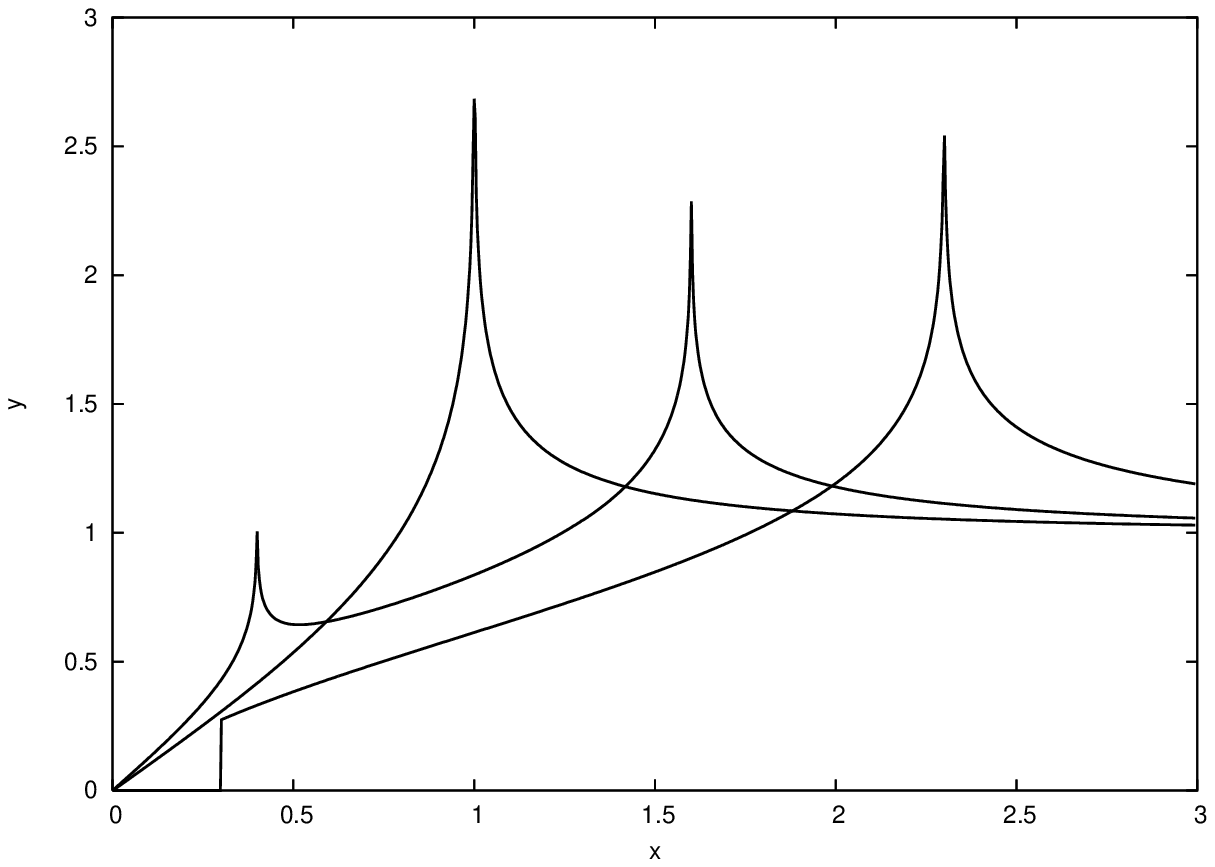}
\caption{\label{f+dos}Left panel: the temperature dependence of the function $h(T;\{D_0,D_2\})$ for fixed $D_2/T_c=0.7$ and for different $D_0/D_2$ values, where $h_0(T)\equiv h(T;\{0,D_2\})$. For the numerical computations $v_Fk_F/T_c=30$ was applied. Right panel: the quasiparticle density of states for $\Delta_0/\Delta_2=0$, 0.6, and 1.3. The curve with the highest peak at $|\omega|=\Delta_2$ belongs to the value 0 (one component UCDW), while the one with symmetrically placed smaller peaks belongs to 0.6, finally the curve with one peak and a clean gap belongs to 1.3.}
\end{figure}

In this subsection we focus on the solutions of the gap equation obtained in Eq.~\eqref{gap-equation}. The knowledge of the explicit momentum and temperature dependence of $\Delta$ is essential for our further investigations. So far the calculation is completely general within the frames of the Hamiltonian in Eq.~\eqref{hamiltonian}. For further study however, we retain only the longitudinal term in the interaction in Eq.~\eqref{interaction}, which is considered to be the strongest component, and ignore the coupling to the transverse phonon modes. Therefore we drop the polarization index $\lambda$ in the followings for brevity. With this, the kernel simplifies to $P(\mathbf{k,k'})=4\omega(\mathbf{Q})^{-1}D^*(\mathbf{k})D(\mathbf{k'})$. The immediate consequence of the factorization of $P$ in its momentum arguments, that follows right from the gap equation and from the explicit form of $D^*(\mathbf{k})$ (see Eq.~\eqref{coupling-expanded}) is
\begin{equation}\label{fraction}
\frac{\Delta_i(T)}{\Delta_j(T)}=\frac{D_i}{D_j},\quad\quad i,j=0,\dots,4.
\end{equation}
The relation in Eq.~\eqref{fraction} not only fixes the ratios of the different gap amplitudes to a given, temperature independent value, but also involves that there is only one $T_c$, where each type of phases (with nonzero coupling) simultaneously open. Accordingly, the solution of Eq.~\eqref{fraction} is simply $\Delta_i(T)=D_{i}h$. Here $h(T;\{D_i\})$ is a dimensionless function of order unity determined by
\begin{equation}
1=\frac{2}{\omega(\mathbf{Q})N}\sum_{\mathbf{k}}\tanh\left(\frac{\sqrt{\xi(\mathbf{k})^2+h^2|D(\mathbf{k})|^2}}{2T}\right)\frac{|D(\mathbf{k})|^2}{\sqrt{\xi(\mathbf{k})^2+h^2|D(\mathbf{k})|^2}},
\end{equation}
from which the transition temperature is obtained as
\begin{equation}\label{transition}
T_c=\frac{2\gamma}{\pi}v_Fk_F e^{-\omega(\mathbf{Q})/[2\rho_0(0)\langle|D|^2\rangle]}.
\end{equation}
In equation~\eqref{transition} $\rho_0(0)=a/\pi v_F$ stands for the normal state density of states per spin, and $\langle\dots\rangle$ denotes the Fermi surface average. Furthermore, at zero temperature and around $T_c$ we have
\begin{align}
h(0)&=\frac{\pi}{\gamma}\exp\left(-\frac{\langle|D|^2\ln(T_c^{-1}|D|)\rangle}{\langle|D|^2\rangle}\right),\\
h(T)&=\pi T_c\sqrt{\frac{8}{7\zeta(3)}}\sqrt{\frac{\langle|D|^2\rangle}{\langle|D|^4\rangle}}\sqrt{1-\frac{T}{T_c}}.
\end{align}
The whole temperature dependence of $h$ below $T_c$ is shown in Fig.~\ref{f+dos} (left panel) for the case where only $D_0$ and $D_2$ are set to nonzero, real and positive values in Eq.~\eqref{coupling-expanded}. This restricted choice of the parameters is sufficient and simple enough to present results for a coexisting CDW+UCDW phase. Consequently, the order parameter will have the form
\begin{equation}\label{coexisting-delta}
\Delta(\mathbf{k})=\Delta_0+\Delta_2\sin(bk_y),
\end{equation}
with the gap amplitudes both positive. At this point it is important to call the attention to the fact, that since the coupling amplitudes $D_i$ in Eq.~\eqref{coupling-expanded} can be complex as well, it is even possible that there is a relative phase between $\Delta_0$ and $\Delta_2$ leading to an optical gap in the single particle density of states regardless of the values of $|\Delta_0|$ and $|\Delta_2|$, but this case is beyond our scope of investigation in this paper. We note however, that complexity of the same order parameter in electronic UCDW can account for a number of peculiar properties observed in (TaSe$_4$)$_2$I.\cite{tase4}

Since $\Delta(\mathbf{k})=h(T)D^*(\mathbf{k})$, thus Eq.~\eqref{order-definition} implies that $\langle\phi\rangle=\sqrt{N}h(T)$ for the longitudinal phonons. Consequently the $\mathbf{Q}$ Fourier component of the lattice distortion can be written as $\mathbf{u(Q)}\simeq g\langle\phi\rangle\,\mathbf{e}(\mathbf{Q})\sim h(T)$. This Peierls distortion of the ions can be seen by x-rays even if the electron system is unconventional.

Now we calculate the quasiparticle density of states for the coexisting phase having the type of gap function in Eq.~\eqref{coexisting-delta}. The density of states reads as
\begin{equation}\label{dos}
\begin{split}
\frac{\rho(\omega)}{\rho_0(0)}&=\int_0^{2\pi}\frac{\text{d}y}{2\pi}
\text{Re}\frac{|\omega|}{\sqrt{\omega^2-|\Delta_0+\Delta_2\sin(y)|^2}}\\
&=\frac{2}{\pi}\text{Re}\left\{
\frac{|\omega|}{\sqrt{(|\omega|+\Delta_2)^2-\Delta_0^2}}K\left(2\sqrt{\frac{|\omega|\Delta_2}{(|\omega|+\Delta_2)^2-\Delta_0^2}}\right)
\right\},
\end{split}
\end{equation}
where $K(z)$ is the complete elliptic integral of the first kind. The energy dependence of $\rho(\omega)$ is shown in Fig.~\ref{f+dos} (right panel) for different values of $\Delta_0/\Delta_2$. It is clear from the figure and can be easily verified from Eq.~\eqref{dos}, that a true optical gap $G=2(\Delta_0-\Delta_2)$ opens at the Fermi energy only if  $\Delta_0>\Delta_2$, furthermore the positions of the logarithmically divergent peaks are given by
\begin{equation}\label{peaks}
|\omega_\text{peak}|=\Theta(\Delta_2-\Delta_0)|\Delta_0\pm\Delta_2|+\Theta(\Delta_0-\Delta_2)(\Delta_0+\Delta_2).
\end{equation}
One can also easily check, that in the $\Delta_0=0$ limit ($D_0=0$), Eq.~\eqref{dos} simplifies to the well known result of a single component unconventional density wave,\cite{balazs-sdw} and obviously the $\Delta_2=0$ limit ($D_2=0$) yields the conventional BCS result.

\section{Correlation functions}

In this section we calculate the frequency dependent conductivity of the previously introduced model, namely that of the coexisting CDW+UCDW phase with $\Delta(\mathbf{k})=\Delta_0+\Delta_2\sin(bk_y)$. The outline is what follows: we shall start with the formulation of the Green's function. We then proceed with the evaluation of the quasiparticle contribution of the density-density correlator. We do this, because we are primarily focused on the optical conductivity in the chain direction, which can be determined from the charge conservation rule, that is the continuity equation. Afterwards we study the one bubble result for the conductivity, we go on and incorporate the effect of interaction between electrons and phonons into the theory on the level of random phase approximation (RPA). The RPA calculation enables us to identify the effective mass of the collective motion, the sliding of the density wave.

\subsection{Single particle conductivity}

The Green's function of a charge density wave, either conventional or any type of unconventional, is given by $G^{-1}(\mathbf{k},i\omega_n)=i\omega_n-H(\mathbf{k})$, where $H(\mathbf{k})=\xi(\mathbf{k})\rho_3+\Delta'(\mathbf{k})\rho_1-\Delta''(\mathbf{k})\rho_2$ is the four by four matrix appearing in the mean field Hamiltonian in Eq.~\eqref{hamiltonian-meanfield}. Here, unconventionality is manifested solely in the momentum dependence of the gap. From this, $G$ is explicitly obtained as
\begin{equation}
G(\mathbf{k},i\omega_n)=-\frac{i\omega_n+\xi(\mathbf{k})\rho_3+\Delta'(\mathbf{k})\rho_1-\Delta''(\mathbf{k})\rho_2}{\omega_n^2+\xi(\mathbf{k})^2+|\Delta(\mathbf{k})|^2}.
\end{equation}
Now, the single particle contribution to the retarded product of the density correlator, $\Pi_{11}^0\equiv\langle[n,n]\rangle_0$, is calculated in the usual way and reads as\cite{balazs-collective}
\begin{equation}\label{density0-corr}
\begin{split}
\Pi^0_{11}(\zeta,i\nu\to\omega)&=-\frac{1}{\beta N}\sum_{\mathbf{k},\omega_n}
\text{Tr}(G(\mathbf{k},i\omega_n)G(\mathbf{k+q},i\omega_n+i\nu))\\
&=2\rho_0(0)\frac{\zeta^2}{\zeta^2-\omega^2}(1-f),
\end{split}
\end{equation}
where, for simplicity, we limited our analysis to $\mathbf{q}=(q_x,0,0)$ (i.e. wave vector pointing in the quasi-one dimensional direction) and $\zeta=v_Fq_x$. In addition, $f$ is the generalized version of the function that also appears in the correlation functions of conventional DWs with constant gap,\cite{viro2} only minor differences are present due to incorporation of unconventionality
\begin{eqnarray}\label{f-function}
f&=&(\zeta^2-\omega^2)\frac{2}{\pi}\int_0^\infty\int_0^{2\pi}\tanh\left(\frac{\beta E}{2}\right)\frac{N}{D}
\text{Re}\frac{|\Delta(y)|^2}{\sqrt{E^2-|\Delta(y)|^2}}\,\text{d}y\,\text{d}E,\\
N&=&(\zeta^2-\omega^2)^2-4E^2(\zeta^2+\omega^2)+4\zeta^2|\Delta(y)|^2,\notag\\
D&=&N^2-64E^2\omega^2\zeta^2(E^2-|\Delta(y)|^2),\notag
\end{eqnarray}
where $\Delta(y)=\Delta_0+\Delta_2\sin(y)$.

\begin{figure}
\psfrag{x}[t][b]{$|\omega|/\Delta_2$}
\psfrag{y}[b][t]{$\sigma_1(\omega)/[ne^2/(m2\Delta_2)]$}
\includegraphics[width=8.2cm,height=6cm]{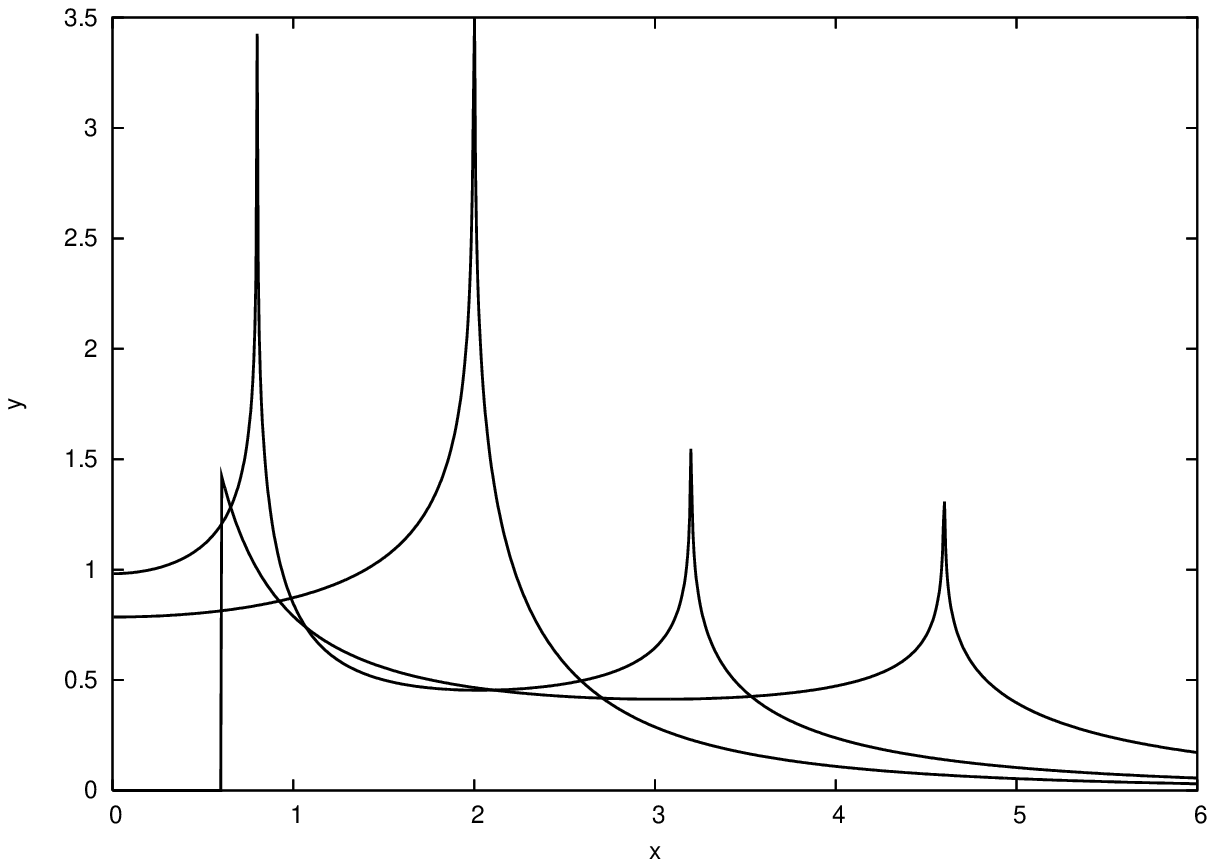}\hfill
\psfrag{y}[b][t]{$\omega\sigma_2(\omega)/[ne^2/m]$}
\includegraphics[width=8.2cm,height=6cm]{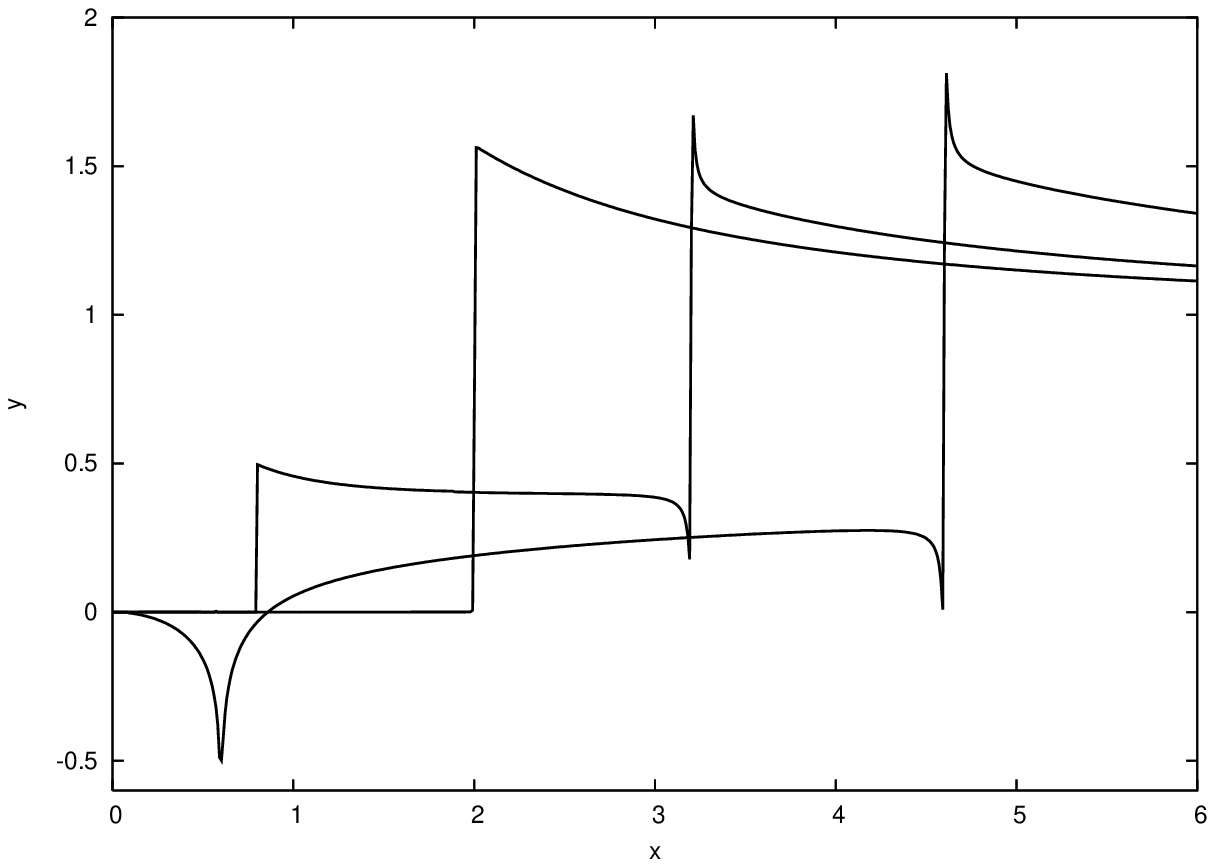}
\caption{\label{sigma}Left panel: the real part of the optical conductivity in the quasi-one dimensional direction at zero temperature for $\Delta_0/\Delta_2=0$, 0.6 and 1.3, respectively. The curve with a single peak at $|\omega|=2\Delta_2$ belongs to the value 0, and is the result for a single component UCDW. The curve with two symmetrically placed peaks around $2\Delta_2$ belongs to 0.6. Finally the spectrum with a clean gap around zero frequency has $\Delta_0=1.3\Delta_2$. Right panel: the imaginary part of the optical conductivity at zero temperature for the same $\Delta_0/\Delta_2$ values that appear on the left panel. The line with a positive jump of $\pi/2$ at $|\omega|=2\Delta_2$ is the single component UCDW result with $\Delta_0=0$. The other two belong to 0.6 and 1.3 (from left to right).}
\end{figure}

Now, making use of Eq.~\eqref{density0-corr} and charge conservation, the single particle contribution to the complex conductivity in the chain direction is of the form $\sigma(\zeta,\omega)=ne^2m^{-1}i\omega(\omega^2-\zeta^2)^{-1}(1-f)$, where $n/m=2\rho_0(0)v_F^2/V_c$, with $n$ being the particle density and $V_c$ is the cell volume. Furthermore, in the $\zeta\to0$ long wavelength limit, the real and imaginary parts of the optical conductivity $(\sigma=\sigma_1+i\sigma_2)$ are given by
\begin{eqnarray}
\sigma_1&=&D\delta(\omega)+(ne^2/m)\omega^{-1}f_0'',\label{real}\\
\sigma_2&=&(ne^2/m)\omega^{-1}(1-f_0'),\label{imag}
\end{eqnarray}
where $f_0\equiv\lim_{\zeta\to0}f=f_0'+if_0''$ is the decomposition of $f_0$ into real and imaginary parts, and $D=(ne^2\pi/m)[1-f_0(0)]$ is the Drude weight. The regular part of $\sigma_1$ in Eq.~\eqref{real} can be calculated explicitly, since $f_0''$ has a closed form if both gap amplitudes are real. That is, we again constrain our analysis to a real gap function, just as we did in the previous section (see Eq.~\eqref{coexisting-delta} and subsequent argument). With all this
\begin{equation}
f_0''(\omega)=\tanh\left(\frac{\omega}{4T}\right)\frac{\text{Re}\{C(x,u)\}}{4x}.
\end{equation}
Here
\begin{multline}
C(x,u)=\frac{1}{\sqrt{(x+1)^2-u^2}}\left\{2[(u+1)^2+x^2]K\left(2\sqrt{\frac{x}{(x+1)^2-u^2}}\right)\right.\\
\left.-2[(x+1)^2-u^2]E\left(2\sqrt{\frac{x}{(x+1)^2-u^2}}\right)
-4u(u-x+1)\Pi\left(\frac{2x}{x+u+1},2\sqrt{\frac{x}{(x+1)^2-u^2}}\right)\right\},
\end{multline}
$x=|\omega|/(2\Delta_2)$, $u=\Delta_0/\Delta_2$, $K(z)$ and $E(z)$ are the complete elliptic integrals of the first and second kind, while $\Pi(n,z)$ stands for the incomplete elliptic integral of the third kind. The imaginary part $\sigma_2$ in Eq.~\eqref{imag} is evaluated numerically with Kramers-Kronig transformation at zero temperature. The frequency dependence of both quantities are shown in Fig.~\ref{sigma} for different values of the ratio $\Delta_0/\Delta_2$, including the case of $\Delta_0=0$, which is simply the one component UCDW limit.\cite{balazs-sdw}
It is interesting to see from Fig.~\ref{sigma} that whatever small the $\Delta_2$ is, while we are in the $\Delta_0>\Delta_2$ regime, the sharp onset of absorption with square root divergence at the gap edge known for a conventional density wave is disappeared. The size of the gap is $G=2(\Delta_0-\Delta_2)$, that has been mentioned already along with the single particle density of states (see Fig.~\ref{f+dos}). In addition, the peaks of logarithmic type are apparently located at $2\omega_\text{peak}$, where $\omega_\text{peak}$ is the corresponding singular point in the quasiparticle density of states defined in Eq.~\eqref{peaks}.

\subsection{Conductivity with collective contribution}

We shall turn now our attention to the effect of collective contributions on the results of the previous subsection. Namely, we investigate how the conductivity is affected if we include the short wavelength component of the interaction within RPA. This part of the interaction is responsible for the density wave instability and depending on the strength of the couplings in Eq.~\eqref{coupling-expanded}, it drives the system into one of the possible symmetry breaking ground states, either a conventional CDW, a single component unconventional CDW, or even one of the many coexisting phases. We proceed with the assumption we made through the paper, that is the ground state of the interacting electron-phonon system is a coexisting CDW+UCDW phase with order parameter $\Delta(\mathbf{k})=\Delta_0+\Delta_2\sin(bk_y)$. Our goal is to calculate the dressed chain direction conductivity by taking into account the fluctuation of the order parameter, leading to the sliding of the condensate as a whole. With this, we are in a position to identify the effective mass of the collective excitation, the same way as it was done in the seminal paper of Lee, Rice and Anderson.\cite{lra}

The interaction part of the Hamiltonian in Eq.~\eqref{hamiltonian}, describing scatterings of the electrons from one Fermi sheet to the other, can be recast as
\begin{equation}\label{interact}
H'=\frac{1}{\sqrt{N}}\sum_{\mathbf{q,k}}\{D(\mathbf{k,-Q})\Psi^+(\mathbf{k+q})\rho_-\Psi(\mathbf{k})\phi^+(\mathbf{Q})
+\text{h.c.}\},
\end{equation}
where $\rho_\pm=(\rho_1\pm i\rho_2)/2$, and only the coupling to the longitudinal phonon has been retained in accordance with what has been pointed out in the subsection dealing with the solutions of the gap equation. Namely, it is a reasonable approximation to omit the coupling to the transverse modes, as they do not really form ionic charge fluctuations that the electrons could feel and couple to.
The bare phonon propagator, that acts as the interaction between electrons in the diagrammatic language, is given by
\begin{equation}
\mathcal{D}^0(\mathbf{q},i\omega_n)=-\int_0^\beta\text{d}\tau\,e^{i\omega_n\tau}\langle T_\tau\phi(\mathbf{q},\tau)\phi(-\mathbf{q})\rangle=\frac{2\omega(\mathbf{q})}{(i\omega_n)^2-\omega(\mathbf{q})^2}.
\end{equation}

\begin{figure}
\psfrag{x}[t][b]{$|\omega|/\Delta_2$}
\psfrag{y}[b][t]{$\sigma_1(\omega)/[ne^2/(m2\Delta_2)]$}
\psfrag{c}[B][B][1][0]{(a)}
\psfrag{1}[B][B][0.8][0]{$\Delta_0/\Delta_2=0$}
\psfrag{4}[B][B][0.8][0]{$0.6$}
\psfrag{6}[B][B][0.8][0]{$1.3$}
\includegraphics[width=8.2cm,height=5.6cm]{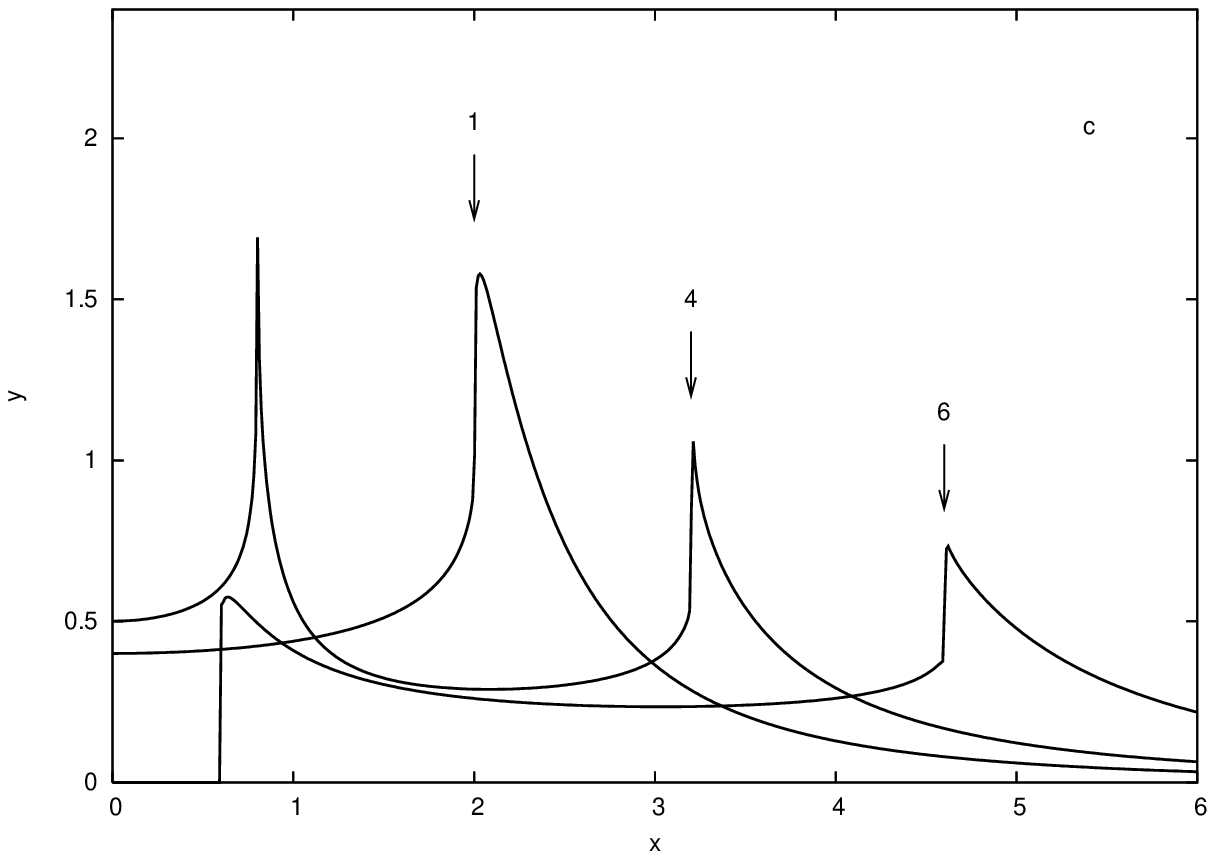}\hfill
\psfrag{d}[B][B][1][0]{(b)}
\psfrag{1}[B][B][0.8][0]{$\Delta_0/\Delta_2=0$}
\psfrag{4}[B][B][0.8][0]{$0.6$}
\psfrag{6}[B][B][0.8][0]{$1.3$}
\psfrag{y}[b][t]{$\omega\sigma_2(\omega)/[ne^2/m]$}
\includegraphics[width=8.2cm,height=5.6cm]{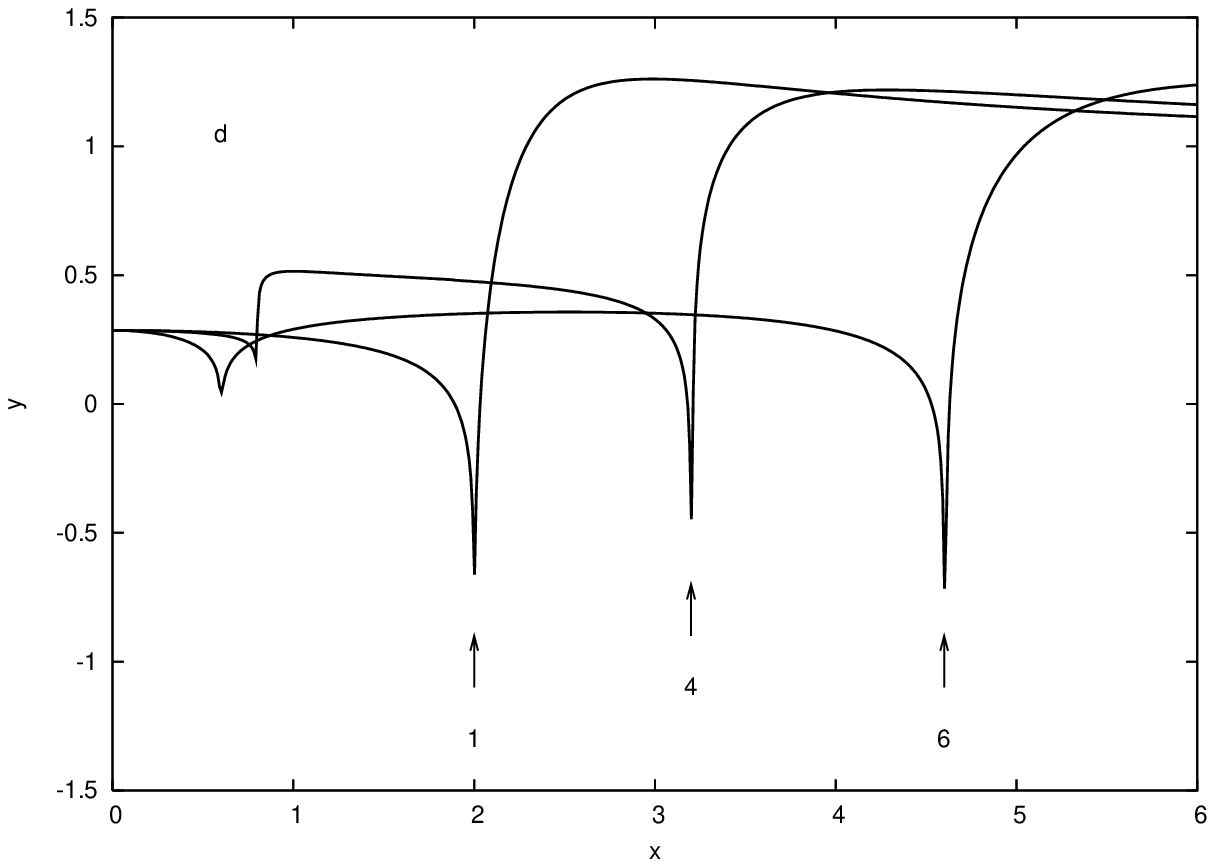}

\medskip
\psfrag{y}[b][t]{$\sigma_1(\omega)/[ne^2/(m2\Delta_2)]$}
\psfrag{e}[B][B][1][0]{(c)}
\includegraphics[width=8.2cm,height=5.6cm]{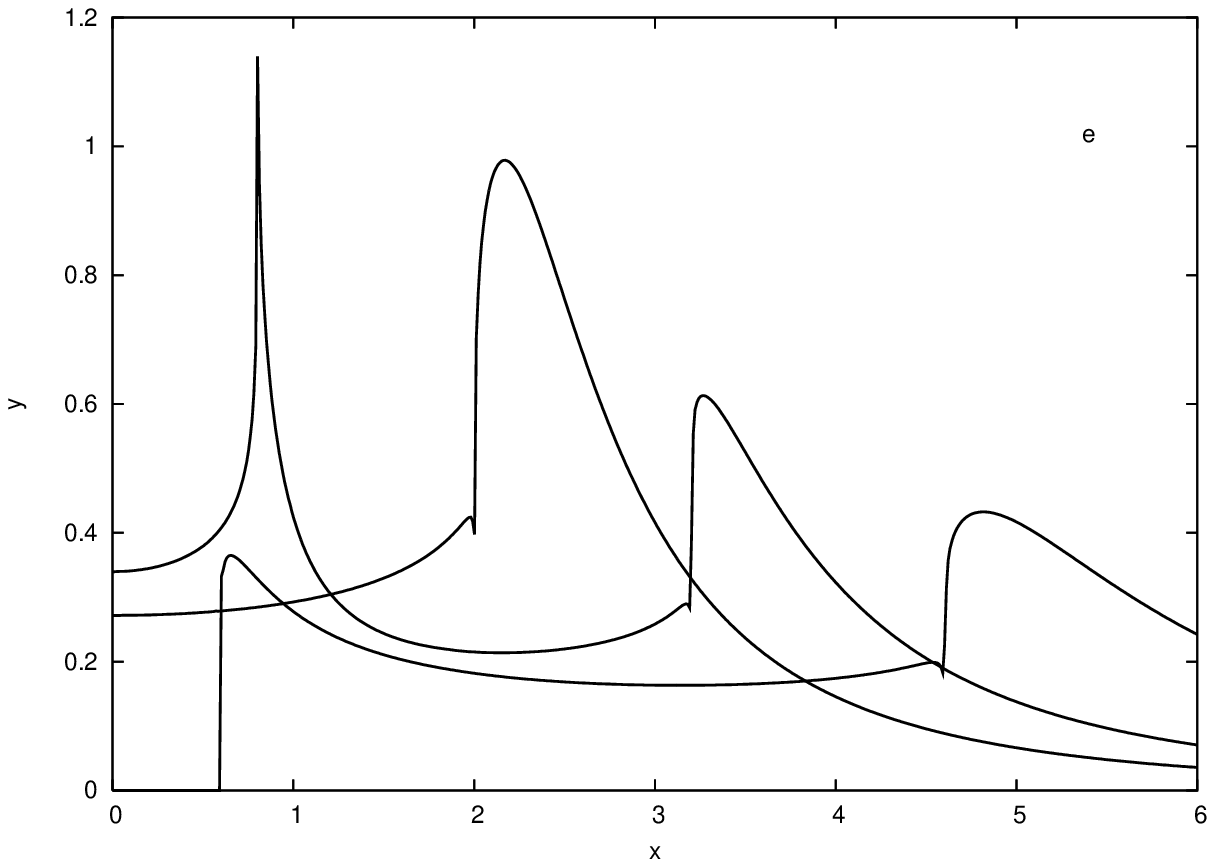}\hfill
\psfrag{f}[B][B][1][0]{(d)}
\psfrag{y}[b][t]{$\omega\sigma_2(\omega)/[ne^2/m]$}
\includegraphics[width=8.2cm,height=5.6cm]{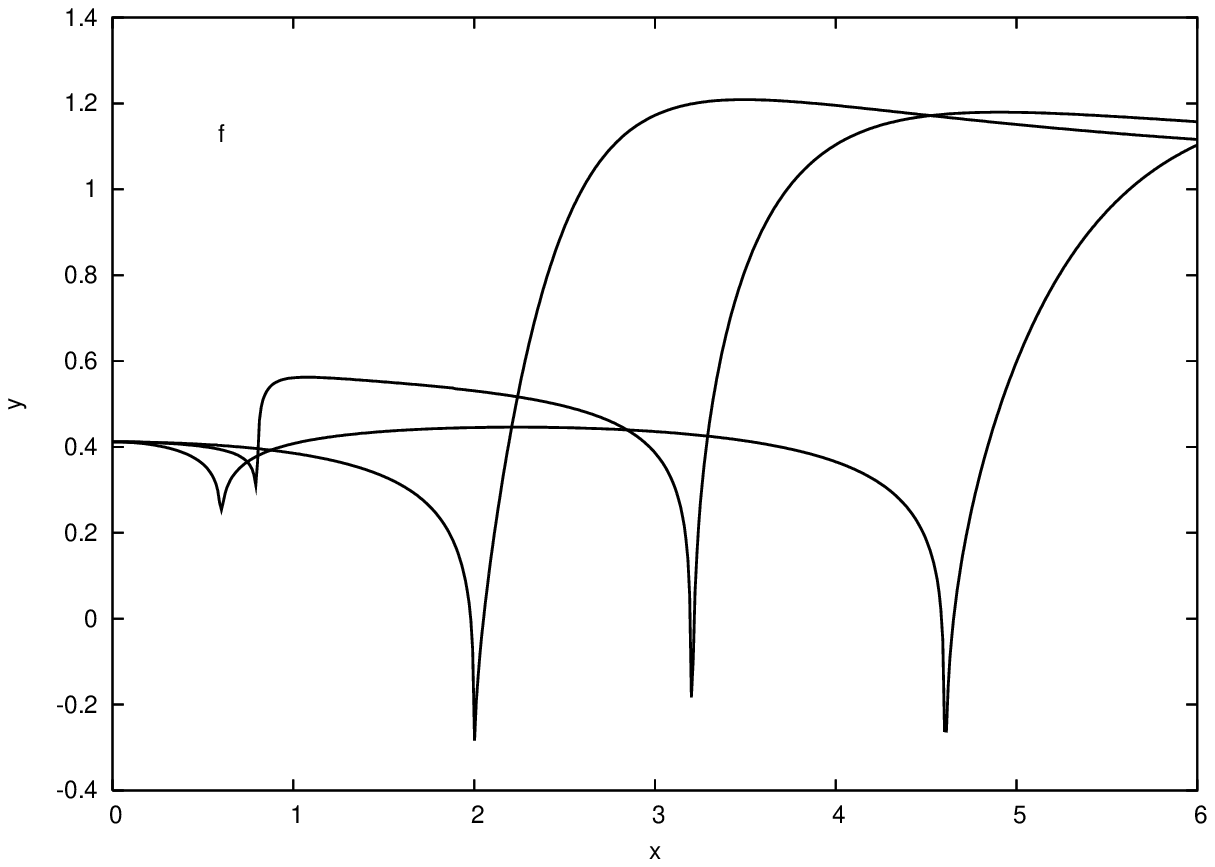}

\medskip
\psfrag{y}[b][t]{$\sigma_1(\omega)/[ne^2/(m2\Delta_2)]$}
\psfrag{g}[B][B][1][0]{(e)}
\includegraphics[width=8.2cm,height=5.6cm]{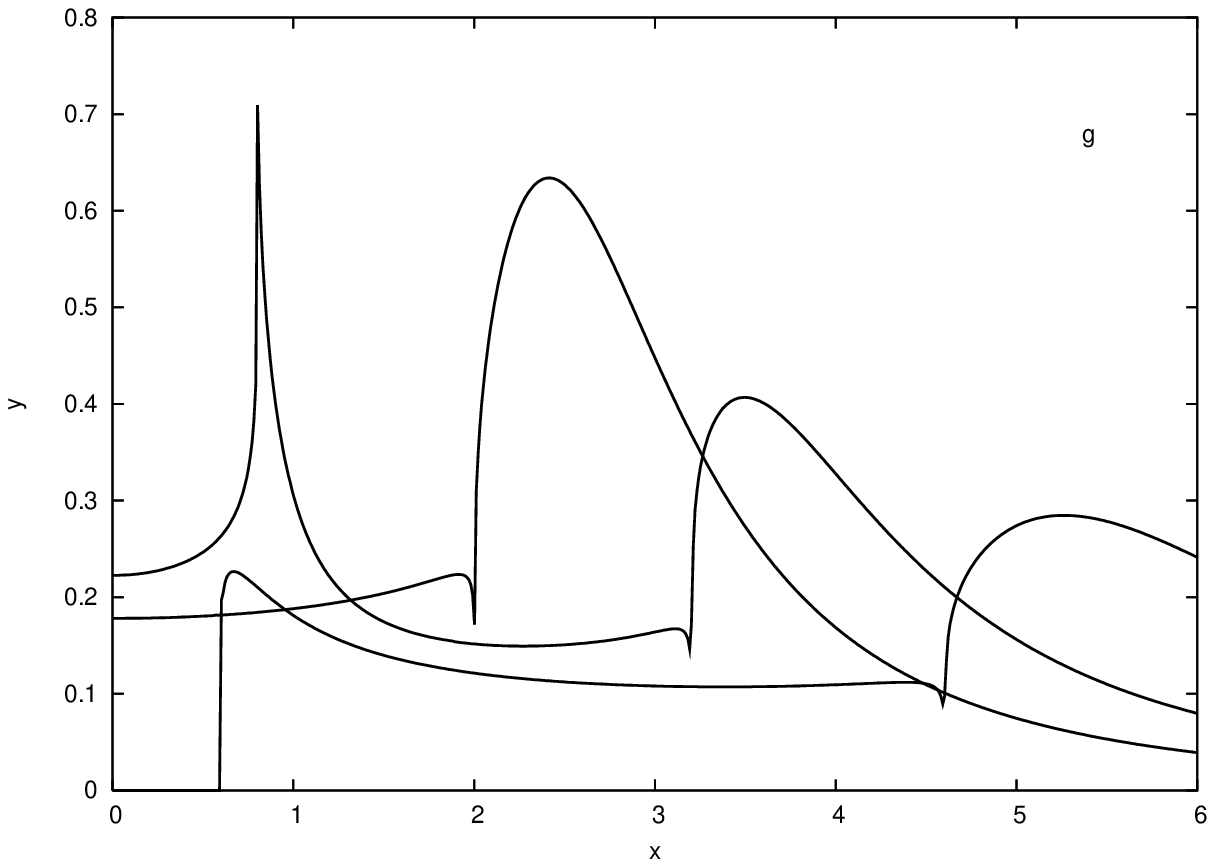}\hfill
\psfrag{y}[b][t]{$\omega\sigma_2(\omega)/[ne^2/m]$}
\psfrag{h}[B][B][1][0]{(f)}
\includegraphics[width=8.2cm,height=5.6cm]{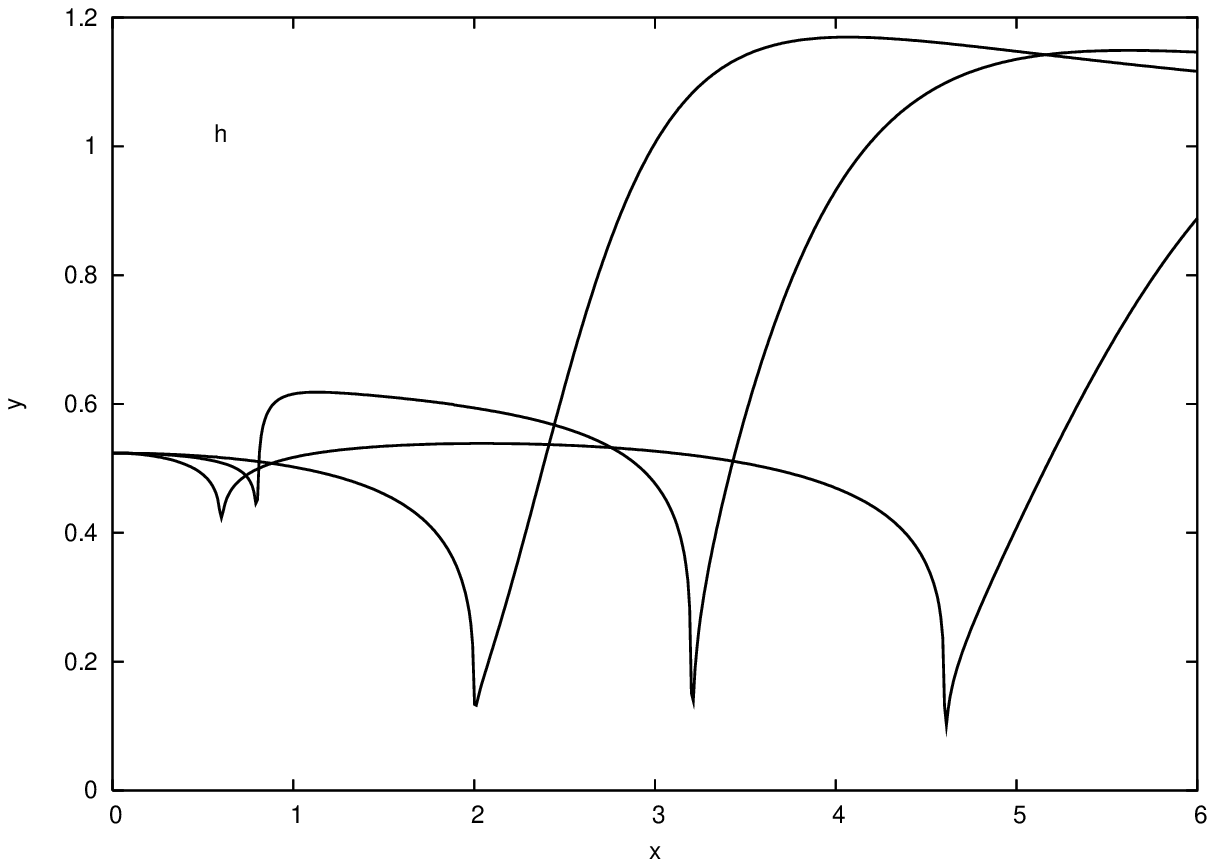}
\caption{\label{sigma-rpa}The real (left panels) and imaginary parts (right panels) of the optical conductivity in the RPA at zero temperature from Eqs.~\eqref{sigma1-rpa} and~\eqref{sigma2-rpa}. The $\Delta_0/\Delta_2$ ratio is fixed to the same values as in Fig.~\ref{sigma} (one bubble results), that are 0, 0.6, and 1.3. Beside that, the dimensionless, renormalized coupling appearing in the RPA formulae is set to $\lambda(0)=0.4$ for figures (a) and (b), $\lambda(0)=0.7$ for (c) and (d), and $\lambda(0)=1.1$ for (e) and (f), respectively.}
\end{figure}

\begin{figure}
\psfrag{x}[t][b]{$T/T_c$}
\psfrag{y}[b][t]{$m^*(T)/m^*(0)$}
\psfrag{1}[Bc][Bc][0.8][0]{$D_0/D_2=0.6$}
\psfrag{2}[Bc][Bc][0.8][0]{0}
\psfrag{3}[Bc][Bc][0.8][0]{conv. CDW}
\psfrag{4}[Bc][Bc][0.8][0]{1.3}
\includegraphics[width=8.2cm,height=6cm]{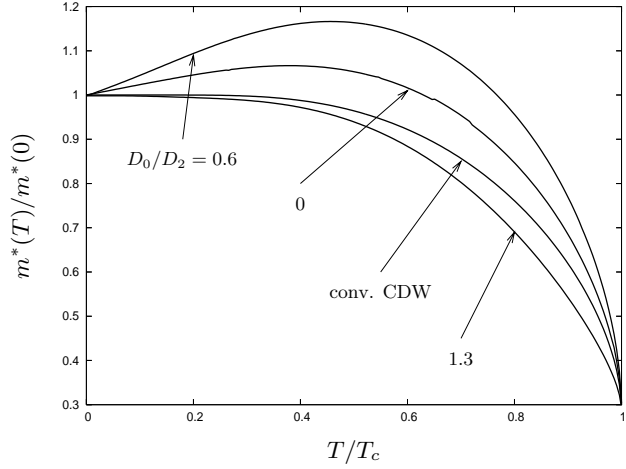}
\caption{\label{mass}. The effective mass versus reduced temperature for both a conventional CDW and a CDW+UCDW phase with $\lambda(0)=0.4$. In the latter case, we present results for $\Delta_0(T)/\Delta_2(T)\equiv D_0/D_2=0$, 0.6, and 1.3 with $D_2/T_c=0.7$ fixed. The CDW result is plotted for comparison, where we used $D_2=0$, and $D_0/T_c=0.7$. For each curve $m^*(0)/m=3.5$.}
\end{figure}

With all these, the coupled RPA equations for the renormalized density-density correlator $\Pi_{11}=\langle[n,n]\rangle$ are
\begin{align}\label{dyson}
\Pi_{11}&=\Pi^0_{11}-\Pi^0_{1+}\mathcal{D}^0\Pi_{-1}-\Pi^0_{1-}\mathcal{D}^0\Pi_{+1},\\
\Pi_{+1}&=\Pi^0_{+1}-\Pi^0_{+-}\mathcal{D}^0\Pi_{+1}-\Pi^0_{++}\mathcal{D}^0\Pi_{-1},\\
\Pi_{-1}&=\Pi^0_{-1}-\Pi^0_{-+}\mathcal{D}^0\Pi_{-1}-\Pi^0_{--}\mathcal{D}^0\Pi_{+1}.
\end{align}
Here $\Pi_{11}^0$ is given by Eq.~\eqref{density0-corr}, and the other zeroth order correlators (with superscript 0) are also evaluated within the standard method, and we find
\begin{equation}
\begin{split}
\Pi^0_{\pm\mp}(\zeta,i\nu\to\omega)&=-\frac{1}{\beta N}\sum_{\mathbf{k},\omega_n}|D(\mathbf{k})|^2\text{Tr}(\rho_\pm G(\mathbf{k},i\omega_n)\rho_\mp G(\mathbf{k+q},i\omega_n+i\nu))\\
&=\frac{\omega(\mathbf{Q})}{2}+\frac{\rho_0(0)}{4h^2}\left[(\omega^2-\zeta^2)f-2f_4\right],
\end{split}
\end{equation}
and similarly
\begin{align}
\Pi^0_{\pm\pm}(\zeta,i\nu\to\omega)&=-\frac{\rho_0(0)}{2h^2}f_4,\\
\Pi^0_{1\pm}(\zeta,i\nu\to\omega)&=\pm\frac{\rho_0(0)}{2h}\zeta f.
\end{align}
In addition, $f_4$ differs from $f$ only in the integrand in Eq.~\eqref{f-function}, namely there is an extra $|\Delta(y)|^2$ factor in the numerator. After some algebra, one finds for the dressed density correlator
\begin{equation}\label{correlator}
\Pi_{11}=2\rho_0(0)\frac{\zeta^2}{\zeta^2-\omega^2}\left(1-\frac{f\omega^2}
{\omega^2+\lambda f(\omega^2-\zeta^2)}\right),
\end{equation}
where $\lambda=\lambda_0[\omega^2(\mathbf{Q})/\langle|2\Delta|^2\rangle]\equiv\omega(\mathbf{Q})\rho_0(0)/(2h^2)$ is the renormalized temperature dependent electron-phonon coupling, while $\lambda_0=2\rho_0(0)\langle|D|^2\rangle/\omega(\mathbf{Q})$ is the bare coupling, see Eq.~\eqref{transition}. With the aid of Eq.~\eqref{correlator} one ends up with the following formula for the inchain optical conductivity
\begin{equation}
\sigma(\omega)=\frac{ne^2}{m}\frac{1}{i\omega}\left(-1+f_0-f_0\frac{m}{m^*}\right),
\end{equation}
where $m^*/m=1+\lambda^{-1}f_0^{-1}$ is the effective mass. The real and imaginary parts ($\sigma=\sigma_1+i\sigma_2$) read as
\begin{align}
\sigma_1&=D\delta(\omega)+\frac{ne^2}{m}\frac{f_0''}{\omega|1+\lambda f_0|^2},\label{sigma1-rpa}\\
\sigma_2&=\frac{ne^2}{m}\frac{1}{\omega}\left(1-\frac{f_0'+\lambda|f_0|^2}{|1+\lambda f_0|^2}\right),\label{sigma2-rpa}
\end{align}
with $D=(ne^2\pi/m)[1-\lambda^{-1}m/m^*]$ the Drude weight. Both $\sigma_1$ and $\sigma_2$ are shown in Fig.~\ref{sigma-rpa} versus frequency at zero temperature, for different values of $\lambda(0)$ and $\Delta_0/\Delta_2$. Now we have to point out one interesting feature of $\sigma_1$, that cannot be seen properly from the figure. As the function $f_0''$ has logarithmic singularities at frequencies $\omega=2\omega_\text{peak}$, furthermore $f_0'$ is bounded, this implies that the regular part of $\sigma_1$ in Eq.~\eqref{sigma1-rpa} vanishes in an inverse logarithmic manner at the same frequencies. Since the logarithmic singularity is very weak, the plots in Fig.~\ref{sigma-rpa} cannot really resolve this kind of behavior around the zeros. Nevertheless, as the coupling $\lambda(0)$ increases, the feature becomes more and more apparent. We shall note at this point, that the way how the RPA result relates to the quasiparticle contribution shown in Fig.~\ref{sigma} can be contrasted to what has been found for a conventional charge density wave.\cite{lra} In such a system, the familiar square root singularity at the gap edge $2\Delta$ is suppressed and transformed to a square root edge by the collective contributions. In our phononic UCDW however, the absorption is finite for the whole energy range, unless $\Delta_0>\Delta_2$, where a clean optical gap develops in the spectrum.

The effective mass of the sliding for our CDW+UCDW model can be seen in Fig.~\ref{mass} with $\lambda(0)=0.4$. The parameters used for the numerical calculations are: $\Delta_0(T)/\Delta_2(T)\equiv D_0/D_2=0$, 0.6, and 1.3, respectively with $D_2/T_c=0.7$ being fixed (see the results for $h(T/T_c)$ on Fig.~\ref{f+dos}, left panel). Along with these new results, we have also plotted the effective mass of a conventional CDW for comparison, where we used $D_0/T_c=0.7$. Looking at the figure, an interesting new feature of the phononic UCDW can be established: as long as $D_0$ is smaller than $D_2$, that is the system possesses small energy excitations, Dirac fermions, around the nodes of $\Delta(\mathbf{k})$, the effective mass $m^*$ is a nonmonotonic function of the temperature. That is not the case for a conventional CDW, as it decreases in a monotonic fashion as we approach $T_c$ from below.\cite{effectivemass}

\section{Conclusions}
In this paper we have developed the mean field theory of a novel type of quasi-one dimensional unconventional charge density wave, where the density wave instability is caused purely by phonons. As the electron-phonon coupling exhibits significant dependence on the momentum of the scattered electron, this leads us to a wavevector dependent order parameter taking different values on different points on the Fermi surface. In particular, it turns out that the explicit form of the single particle gap is identical to that in an electronic UCDW, where in contrary, the phonons do not play any role in the development of the density wave condensate, leaving the lattice unaffected. It is important to realize, that if the system favours an unconventional ordering in the ground state, then due to the vanishing momentum average of the order parameter the system lacks any periodic oscillation in the electronic charge. Though the two type of condensates have apparently much in common, like thermodynamics determined by the Dirac electrons around the nodes of the gap for instance, there is an important difference that has to be emphasized. That is in a phononic UCDW, while the electronic density remains homogeneous, the underlying ionic lattice undergoes a distortion as we enter the low temperature phase, clearly signalling the phase transition. For this reason we might say that the "hidden order" reveals itself in such a way, and could be experimentally accessible and measurable by x-ray scattering.

On the other hand, the aforementioned Peierls distortion is a key feature of the conventional charge density waves arising from the coupling of the lattice to the electronic degrees of freedom. Hence, we conclude that the present model of the phononic UCDW is a natural generalization of the conventional theory to more complicated, even $\mathbf{k}$-dependent couplings. 

We have calculated the optical conductivity in the quasi-one dimensional direction, and identified the effective mass of the collective phase excitation. The calculations were carried out using a multicomponent gap allowing us to investigate the effect of the coexisting conventional and unconventional orders. Consequently, the spectra are more structured compared to the pioneer result of Lee, Rice and Anderson.\cite{lra} Nevertheless, though the singularities and zeros are of logarithmic type characteristic to UDWs in general, the single component UCDW limit shows the familiar property: the singularity at the maximum optical gap is suppressed by the collective contributions. In addition there is considerable absorption for small frequencies arising from the nodal excitations. We have also found that the effective mass exhibits a novel type of nonmonotonic temperature dependence as long as the unconventional order dominates the low temperature phase, as opposed to conventional systems. This could serve as a valuable tool in identifying the nature of the low temperature phase.

\begin{acknowledgments}
This work was supported by the Hungarian National Research Fund under grant numbers OTKA NDF45172, T046269, TS049881 and the Magyary Zolt\'an postdoctoral program of Foundation for Hungarian Higher Education and Research (AMFK).
\end{acknowledgments}

\bibliography{eph}

\begin{thebibliography}{40}
\expandafter\ifx\csname natexlab\endcsname\relax\def\natexlab#1{#1}\fi
\expandafter\ifx\csname bibnamefont\endcsname\relax
  \def\bibnamefont#1{#1}\fi
\expandafter\ifx\csname bibfnamefont\endcsname\relax
  \def\bibfnamefont#1{#1}\fi
\expandafter\ifx\csname citenamefont\endcsname\relax
  \def\citenamefont#1{#1}\fi
\expandafter\ifx\csname url\endcsname\relax
  \def\url#1{\texttt{#1}}\fi
\expandafter\ifx\csname urlprefix\endcsname\relax\def\urlprefix{URL }\fi
\providecommand{\bibinfo}[2]{#2}
\providecommand{\eprint}[2][]{\url{#2}}

\bibitem[{\citenamefont{Gr\"uner}(1994)}]{gruner-book}
\bibinfo{author}{\bibfnamefont{G.}~\bibnamefont{Gr\"uner}},
  \emph{\bibinfo{title}{Density waves in solids}}
  (\bibinfo{publisher}{Addison-Wesley}, \bibinfo{address}{Reading},
  \bibinfo{year}{1994}).

\bibitem[{\citenamefont{Anderson and Morel}(1961)}]{anderson}
\bibinfo{author}{\bibfnamefont{P.~W.} \bibnamefont{Anderson}} \bibnamefont{and}
  \bibinfo{author}{\bibfnamefont{P.}~\bibnamefont{Morel}},
  \bibinfo{journal}{Phys. Rev.} \textbf{\bibinfo{volume}{123}},
  \bibinfo{pages}{1911} (\bibinfo{year}{1961}).

\bibitem[{\citenamefont{Won and Maki}(1999)}]{won-maki}
\bibinfo{author}{\bibfnamefont{H.}~\bibnamefont{Won}} \bibnamefont{and}
  \bibinfo{author}{\bibfnamefont{K.}~\bibnamefont{Maki}},
  \emph{\bibinfo{title}{Symmetry and Pairing in Superconductors}}
  (\bibinfo{publisher}{edited by M. Ausloos and S. Kruchinin},
  \bibinfo{address}{Kluver, Dordrecht}, \bibinfo{year}{1999}).

\bibitem[{\citenamefont{van Harlingen}(1995)}]{harlingen}
\bibinfo{author}{\bibfnamefont{D.~J.} \bibnamefont{van Harlingen}},
  \bibinfo{journal}{Rev. Mod. Phys.} \textbf{\bibinfo{volume}{67}},
  \bibinfo{pages}{515} (\bibinfo{year}{1995}).

\bibitem[{\citenamefont{Tsuei and Kirtley}(2000)}]{tsuei}
\bibinfo{author}{\bibfnamefont{C.~C.} \bibnamefont{Tsuei}} \bibnamefont{and}
  \bibinfo{author}{\bibfnamefont{J.~R.} \bibnamefont{Kirtley}},
  \bibinfo{journal}{Rev. Mod. Phys.} \textbf{\bibinfo{volume}{72}},
  \bibinfo{pages}{969} (\bibinfo{year}{2000}).

\bibitem[{\citenamefont{Sigrist and Ueda}(1991)}]{sigrist}
\bibinfo{author}{\bibfnamefont{M.}~\bibnamefont{Sigrist}} \bibnamefont{and}
  \bibinfo{author}{\bibfnamefont{K.}~\bibnamefont{Ueda}},
  \bibinfo{journal}{Rev. Mod. Phys.} \textbf{\bibinfo{volume}{63}},
  \bibinfo{pages}{239} (\bibinfo{year}{1991}).

\bibitem[{\citenamefont{Halperin and Rice}(1968)}]{halperin}
\bibinfo{author}{\bibfnamefont{B.~I.} \bibnamefont{Halperin}} \bibnamefont{and}
  \bibinfo{author}{\bibfnamefont{T.~M.} \bibnamefont{Rice}},
  \emph{\bibinfo{title}{Solid State Physics}} (\bibinfo{publisher}{edited by F.
  Seitz, D. Turnbull, H. Ehrenreich}, \bibinfo{address}{Academic Press, New
  York}, \bibinfo{year}{1968}).

\bibitem[{\citenamefont{Gulacsi and Gulacsi}(1987)}]{gulacsi}
\bibinfo{author}{\bibfnamefont{Z.}~\bibnamefont{Gulacsi}} \bibnamefont{and}
  \bibinfo{author}{\bibfnamefont{M.}~\bibnamefont{Gulacsi}},
  \bibinfo{journal}{Phys. Rev. B} \textbf{\bibinfo{volume}{36}},
  \bibinfo{pages}{699} (\bibinfo{year}{1987}).

\bibitem[{\citenamefont{D\'ora and Virosztek}(2001)}]{balazs-sdw}
\bibinfo{author}{\bibfnamefont{B.}~\bibnamefont{D\'ora}} \bibnamefont{and}
  \bibinfo{author}{\bibfnamefont{A.}~\bibnamefont{Virosztek}},
  \bibinfo{journal}{Eur. Phys. J. B} \textbf{\bibinfo{volume}{22}},
  \bibinfo{pages}{167} (\bibinfo{year}{2001}).

\bibitem[{\citenamefont{Nayak}(2000)}]{nayak-solo}
\bibinfo{author}{\bibfnamefont{C.}~\bibnamefont{Nayak}},
  \bibinfo{journal}{Phys. Rev. B} \textbf{\bibinfo{volume}{62}},
  \bibinfo{pages}{4880} (\bibinfo{year}{2000}).

\bibitem[{\citenamefont{Benfatto et~al.}(2000)\citenamefont{Benfatto, Caprara,
  and Castro}}]{benfatto}
\bibinfo{author}{\bibfnamefont{L.}~\bibnamefont{Benfatto}},
  \bibinfo{author}{\bibfnamefont{S.}~\bibnamefont{Caprara}}, \bibnamefont{and}
  \bibinfo{author}{\bibfnamefont{C.~D.} \bibnamefont{Castro}},
  \bibinfo{journal}{Eur. Phys. J. B} \textbf{\bibinfo{volume}{17}},
  \bibinfo{pages}{95} (\bibinfo{year}{2000}).

\bibitem[{\citenamefont{Chakravarty et~al.}(2001)\citenamefont{Chakravarty,
  Laughlin, Morr, and Nayak}}]{laughlin}
\bibinfo{author}{\bibfnamefont{S.}~\bibnamefont{Chakravarty}},
  \bibinfo{author}{\bibfnamefont{R.~B.} \bibnamefont{Laughlin}},
  \bibinfo{author}{\bibfnamefont{D.~K.} \bibnamefont{Morr}}, \bibnamefont{and}
  \bibinfo{author}{\bibfnamefont{C.}~\bibnamefont{Nayak}},
  \bibinfo{journal}{Phys. Rev. B} \textbf{\bibinfo{volume}{63}},
  \bibinfo{pages}{094503} (\bibinfo{year}{2001}).

\bibitem[{\citenamefont{Ozaki}(1992)}]{ozaki}
\bibinfo{author}{\bibfnamefont{M.~A.} \bibnamefont{Ozaki}},
  \bibinfo{journal}{Int. J. Quantum Chem.} \textbf{\bibinfo{volume}{42}},
  \bibinfo{pages}{55} (\bibinfo{year}{1992}).

\bibitem[{\citenamefont{{Castro Neto}}(2001)}]{castro-neto}
\bibinfo{author}{\bibfnamefont{A.~H.} \bibnamefont{{Castro Neto}}},
  \bibinfo{journal}{Phys. Rev. Lett.} \textbf{\bibinfo{volume}{86}},
  \bibinfo{pages}{4382} (\bibinfo{year}{2001}).

\bibitem[{\citenamefont{Ikeda and Ohashi}(1998)}]{ikeda}
\bibinfo{author}{\bibfnamefont{H.}~\bibnamefont{Ikeda}} \bibnamefont{and}
  \bibinfo{author}{\bibfnamefont{Y.}~\bibnamefont{Ohashi}},
  \bibinfo{journal}{Phys. Rev. Lett.} \textbf{\bibinfo{volume}{81}},
  \bibinfo{pages}{3723} (\bibinfo{year}{1998}).

\bibitem[{\citenamefont{Ohashi}(1999)}]{ohashi}
\bibinfo{author}{\bibfnamefont{Y.}~\bibnamefont{Ohashi}},
  \bibinfo{journal}{Phys. Rev. B} \textbf{\bibinfo{volume}{60}},
  \bibinfo{pages}{15388} (\bibinfo{year}{1999}).

\bibitem[{\citenamefont{Virosztek et~al.}(2002)\citenamefont{Virosztek, Maki,
  and D\'ora}}]{attila}
\bibinfo{author}{\bibfnamefont{A.}~\bibnamefont{Virosztek}},
  \bibinfo{author}{\bibfnamefont{K.}~\bibnamefont{Maki}}, \bibnamefont{and}
  \bibinfo{author}{\bibfnamefont{B.}~\bibnamefont{D\'ora}},
  \bibinfo{journal}{Int. J. Mod. Phys.} \textbf{\bibinfo{volume}{16}},
  \bibinfo{pages}{1667} (\bibinfo{year}{2002}).

\bibitem[{\citenamefont{N\'emeth et~al.}(2001)\citenamefont{N\'emeth, Matus,
  Kriza, and Alavi}}]{nemeth}
\bibinfo{author}{\bibfnamefont{L.}~\bibnamefont{N\'emeth}},
  \bibinfo{author}{\bibfnamefont{P.}~\bibnamefont{Matus}},
  \bibinfo{author}{\bibfnamefont{G.}~\bibnamefont{Kriza}}, \bibnamefont{and}
  \bibinfo{author}{\bibfnamefont{B.}~\bibnamefont{Alavi}},
  \bibinfo{journal}{Synth. Metals} \textbf{\bibinfo{volume}{120}},
  \bibinfo{pages}{1007} (\bibinfo{year}{2001}).

\bibitem[{\citenamefont{D\'ora et~al.}()\citenamefont{D\'ora, V\'anyolos, and
  Virosztek}}]{tase4}
\bibinfo{author}{\bibfnamefont{B.}~\bibnamefont{D\'ora}},
  \bibinfo{author}{\bibfnamefont{A.}~\bibnamefont{V\'anyolos}},
  \bibnamefont{and}
  \bibinfo{author}{\bibfnamefont{A.}~\bibnamefont{Virosztek}},
  \bibinfo{note}{cond-mat/0511576}.

\bibitem[{\citenamefont{Andres et~al.}(2001)\citenamefont{Andres, Kartsovnik,
  Biberacher, Weiss, Balthes, M\"uller, and Kushch}}]{andres}
\bibinfo{author}{\bibfnamefont{D.}~\bibnamefont{Andres}},
  \bibinfo{author}{\bibfnamefont{M.~V.} \bibnamefont{Kartsovnik}},
  \bibinfo{author}{\bibfnamefont{W.}~\bibnamefont{Biberacher}},
  \bibinfo{author}{\bibfnamefont{H.}~\bibnamefont{Weiss}},
  \bibinfo{author}{\bibfnamefont{E.}~\bibnamefont{Balthes}},
  \bibinfo{author}{\bibfnamefont{H.}~\bibnamefont{M\"uller}}, \bibnamefont{and}
  \bibinfo{author}{\bibfnamefont{N.}~\bibnamefont{Kushch}},
  \bibinfo{journal}{Phys. Rev. B} \textbf{\bibinfo{volume}{64}},
  \bibinfo{pages}{161104(R)} (\bibinfo{year}{2001}).

\bibitem[{\citenamefont{Mori et~al.}(1990)\citenamefont{Mori, Tanaka, Oshima,
  Saito, Mori, Maruyama, and Inokuchi}}]{mori}
\bibinfo{author}{\bibfnamefont{H.}~\bibnamefont{Mori}},
  \bibinfo{author}{\bibfnamefont{S.}~\bibnamefont{Tanaka}},
  \bibinfo{author}{\bibfnamefont{M.}~\bibnamefont{Oshima}},
  \bibinfo{author}{\bibfnamefont{G.}~\bibnamefont{Saito}},
  \bibinfo{author}{\bibfnamefont{T.}~\bibnamefont{Mori}},
  \bibinfo{author}{\bibfnamefont{Y.}~\bibnamefont{Maruyama}}, \bibnamefont{and}
  \bibinfo{author}{\bibfnamefont{H.}~\bibnamefont{Inokuchi}},
  \bibinfo{journal}{Bull. Chem. Soc. Jpn.} \textbf{\bibinfo{volume}{63}},
  \bibinfo{pages}{2138} (\bibinfo{year}{1990}).

\bibitem[{\citenamefont{Kartsovnik et~al.}(1995)\citenamefont{Kartsovnik,
  Kovalev, Laukhin, Schegolev, Ito, Ishiguro, Kushch, Mori, and
  Saito}}]{kartsovnik}
\bibinfo{author}{\bibfnamefont{M.~V.} \bibnamefont{Kartsovnik}},
  \bibinfo{author}{\bibfnamefont{A.~E.} \bibnamefont{Kovalev}},
  \bibinfo{author}{\bibfnamefont{V.~N.} \bibnamefont{Laukhin}},
  \bibinfo{author}{\bibfnamefont{I.~F.} \bibnamefont{Schegolev}},
  \bibinfo{author}{\bibfnamefont{H.}~\bibnamefont{Ito}},
  \bibinfo{author}{\bibfnamefont{T.}~\bibnamefont{Ishiguro}},
  \bibinfo{author}{\bibfnamefont{N.~D.} \bibnamefont{Kushch}},
  \bibinfo{author}{\bibfnamefont{H.}~\bibnamefont{Mori}}, \bibnamefont{and}
  \bibinfo{author}{\bibfnamefont{G.}~\bibnamefont{Saito}},
  \bibinfo{journal}{Synth. Metals} \textbf{\bibinfo{volume}{70}},
  \bibinfo{pages}{811} (\bibinfo{year}{1995}).

\bibitem[{\citenamefont{Basleti\'c et~al.}(2001)\citenamefont{Basleti\'c,
  Korin-Hamzi\'c, Kartsovnik, and M\"uller}}]{basletic}
\bibinfo{author}{\bibfnamefont{M.}~\bibnamefont{Basleti\'c}},
  \bibinfo{author}{\bibfnamefont{B.}~\bibnamefont{Korin-Hamzi\'c}},
  \bibinfo{author}{\bibfnamefont{M.~V.} \bibnamefont{Kartsovnik}},
  \bibnamefont{and} \bibinfo{author}{\bibfnamefont{H.}~\bibnamefont{M\"uller}},
  \bibinfo{journal}{Synth. Metals} \textbf{\bibinfo{volume}{120}},
  \bibinfo{pages}{1021} (\bibinfo{year}{2001}).

\bibitem[{\citenamefont{Fujita et~al.}(2001)\citenamefont{Fujita, Sasaki,
  Yoneyama, Kobayashi, and Fukase}}]{fujita}
\bibinfo{author}{\bibfnamefont{T.}~\bibnamefont{Fujita}},
  \bibinfo{author}{\bibfnamefont{T.}~\bibnamefont{Sasaki}},
  \bibinfo{author}{\bibfnamefont{N.}~\bibnamefont{Yoneyama}},
  \bibinfo{author}{\bibfnamefont{N.}~\bibnamefont{Kobayashi}},
  \bibnamefont{and} \bibinfo{author}{\bibfnamefont{T.}~\bibnamefont{Fukase}},
  \bibinfo{journal}{Synth. Metals} \textbf{\bibinfo{volume}{120}},
  \bibinfo{pages}{1077} (\bibinfo{year}{2001}).

\bibitem[{\citenamefont{Leylekian et~al.}(2003)\citenamefont{Leylekian, Ravy,
  and Pouget}}]{pouget}
\bibinfo{author}{\bibfnamefont{P.~F.} \bibnamefont{Leylekian}},
  \bibinfo{author}{\bibfnamefont{S.}~\bibnamefont{Ravy}}, \bibnamefont{and}
  \bibinfo{author}{\bibfnamefont{J.~P.} \bibnamefont{Pouget}},
  \bibinfo{journal}{Synth. Metals.} \textbf{\bibinfo{volume}{137}},
  \bibinfo{pages}{1271} (\bibinfo{year}{2003}).

\bibitem[{\citenamefont{D\'ora et~al.}(2001)\citenamefont{D\'ora, Virosztek,
  and Maki}}]{balazs-threshold}
\bibinfo{author}{\bibfnamefont{B.}~\bibnamefont{D\'ora}},
  \bibinfo{author}{\bibfnamefont{A.}~\bibnamefont{Virosztek}},
  \bibnamefont{and} \bibinfo{author}{\bibfnamefont{K.}~\bibnamefont{Maki}},
  \bibinfo{journal}{Phys. Rev. B} \textbf{\bibinfo{volume}{64}},
  \bibinfo{pages}{041101(R)} (\bibinfo{year}{2001}).

\bibitem[{\citenamefont{D\'ora et~al.}(2002{\natexlab{a}})\citenamefont{D\'ora,
  Virosztek, and Maki}}]{balazs-magneticthreshold}
\bibinfo{author}{\bibfnamefont{B.}~\bibnamefont{D\'ora}},
  \bibinfo{author}{\bibfnamefont{A.}~\bibnamefont{Virosztek}},
  \bibnamefont{and} \bibinfo{author}{\bibfnamefont{K.}~\bibnamefont{Maki}},
  \bibinfo{journal}{Phys. Rev. B} \textbf{\bibinfo{volume}{65}},
  \bibinfo{pages}{155119} (\bibinfo{year}{2002}{\natexlab{a}}).

\bibitem[{\citenamefont{D\'ora et~al.}(2002{\natexlab{b}})\citenamefont{D\'ora,
  Maki, and Virosztek}}]{balazs-imperfect}
\bibinfo{author}{\bibfnamefont{B.}~\bibnamefont{D\'ora}},
  \bibinfo{author}{\bibfnamefont{K.}~\bibnamefont{Maki}}, \bibnamefont{and}
  \bibinfo{author}{\bibfnamefont{A.}~\bibnamefont{Virosztek}},
  \bibinfo{journal}{Phys. Rev. B} \textbf{\bibinfo{volume}{66}},
  \bibinfo{pages}{165116} (\bibinfo{year}{2002}{\natexlab{b}}).

\bibitem[{\citenamefont{Maki et~al.}(2003)\citenamefont{Maki, D\'ora,
  Kartsovnik, Virosztek, Korin-Hamzi\'c, and Basleti\'c}}]{balazs-ucdw-et}
\bibinfo{author}{\bibfnamefont{K.}~\bibnamefont{Maki}},
  \bibinfo{author}{\bibfnamefont{B.}~\bibnamefont{D\'ora}},
  \bibinfo{author}{\bibfnamefont{M.~V.} \bibnamefont{Kartsovnik}},
  \bibinfo{author}{\bibfnamefont{A.}~\bibnamefont{Virosztek}},
  \bibinfo{author}{\bibfnamefont{B.}~\bibnamefont{Korin-Hamzi\'c}},
  \bibnamefont{and}
  \bibinfo{author}{\bibfnamefont{M.}~\bibnamefont{Basleti\'c}},
  \bibinfo{journal}{Phys. Rev. Lett.} \textbf{\bibinfo{volume}{90}},
  \bibinfo{pages}{256402} (\bibinfo{year}{2003}).

\bibitem[{\citenamefont{D\'ora et~al.}(2002{\natexlab{c}})\citenamefont{D\'ora,
  Maki, Korin-Hamzi\'c, Basleti\'c, Virosztek, Kartsovnik, and
  M\"uller}}]{balazs-eurlett}
\bibinfo{author}{\bibfnamefont{B.}~\bibnamefont{D\'ora}},
  \bibinfo{author}{\bibfnamefont{K.}~\bibnamefont{Maki}},
  \bibinfo{author}{\bibfnamefont{B.}~\bibnamefont{Korin-Hamzi\'c}},
  \bibinfo{author}{\bibfnamefont{M.}~\bibnamefont{Basleti\'c}},
  \bibinfo{author}{\bibfnamefont{A.}~\bibnamefont{Virosztek}},
  \bibinfo{author}{\bibfnamefont{M.~V.} \bibnamefont{Kartsovnik}},
  \bibnamefont{and} \bibinfo{author}{\bibfnamefont{H.}~\bibnamefont{M\"uller}},
  \bibinfo{journal}{Europhys. Lett.} \textbf{\bibinfo{volume}{60}},
  \bibinfo{pages}{737} (\bibinfo{year}{2002}{\natexlab{c}}).

\bibitem[{\citenamefont{D\'ora et~al.}(2003)\citenamefont{D\'ora, Maki,
  V\'anyolos, and Virosztek}}]{balazs-ucdw-mtpnernst}
\bibinfo{author}{\bibfnamefont{B.}~\bibnamefont{D\'ora}},
  \bibinfo{author}{\bibfnamefont{K.}~\bibnamefont{Maki}},
  \bibinfo{author}{\bibfnamefont{A.}~\bibnamefont{V\'anyolos}},
  \bibnamefont{and}
  \bibinfo{author}{\bibfnamefont{A.}~\bibnamefont{Virosztek}},
  \bibinfo{journal}{Phys. Rev. B} \textbf{\bibinfo{volume}{68}},
  \bibinfo{pages}{241102(R)} (\bibinfo{year}{2003}).

\bibitem[{\citenamefont{V\'anyolos and Virosztek}()}]{ecrys}
\bibinfo{author}{\bibfnamefont{A.}~\bibnamefont{V\'anyolos}} \bibnamefont{and}
  \bibinfo{author}{\bibfnamefont{A.}~\bibnamefont{Virosztek}},
  \bibinfo{note}{accepted in J. Phys. IV.}

\bibitem[{\citenamefont{Fr\"ohlich}(1954)}]{frohlich}
\bibinfo{author}{\bibfnamefont{H.}~\bibnamefont{Fr\"ohlich}},
  \bibinfo{journal}{Proc. R. Soc., London} \textbf{\bibinfo{volume}{A223}},
  \bibinfo{pages}{296} (\bibinfo{year}{1954}).

\bibitem[{\citenamefont{Lee et~al.}(1974)\citenamefont{Lee, Rice, and
  Anderson}}]{lra}
\bibinfo{author}{\bibfnamefont{P.~A.} \bibnamefont{Lee}},
  \bibinfo{author}{\bibfnamefont{T.~M.} \bibnamefont{Rice}}, \bibnamefont{and}
  \bibinfo{author}{\bibfnamefont{P.~W.} \bibnamefont{Anderson}},
  \bibinfo{journal}{Solid State Commun.} \textbf{\bibinfo{volume}{14}},
  \bibinfo{pages}{703} (\bibinfo{year}{1974}).

\bibitem[{\citenamefont{Maki}(1998)}]{maki-dwave}
\bibinfo{author}{\bibfnamefont{K.}~\bibnamefont{Maki}}, \bibinfo{journal}{AIP
  Conf. Proc.} \textbf{\bibinfo{volume}{438}}, \bibinfo{pages}{83}
  (\bibinfo{year}{1998}).

\bibitem[{\citenamefont{Won et~al.}(2005)\citenamefont{Won, Haas, Parker,
  Telang, V\'anyolos, and Maki}}]{maki-nodal}
\bibinfo{author}{\bibfnamefont{H.}~\bibnamefont{Won}},
  \bibinfo{author}{\bibfnamefont{S.}~\bibnamefont{Haas}},
  \bibinfo{author}{\bibfnamefont{D.}~\bibnamefont{Parker}},
  \bibinfo{author}{\bibfnamefont{S.}~\bibnamefont{Telang}},
  \bibinfo{author}{\bibfnamefont{A.}~\bibnamefont{V\'anyolos}},
  \bibnamefont{and} \bibinfo{author}{\bibfnamefont{K.}~\bibnamefont{Maki}},
  \bibinfo{journal}{AIP Conf. Proc.} \textbf{\bibinfo{volume}{789}},
  \bibinfo{pages}{3} (\bibinfo{year}{2005}).

\bibitem[{\citenamefont{D\'ora et~al.}(2004)\citenamefont{D\'ora, Maki, and
  Virosztek}}]{balazs-modern}
\bibinfo{author}{\bibfnamefont{B.}~\bibnamefont{D\'ora}},
  \bibinfo{author}{\bibfnamefont{K.}~\bibnamefont{Maki}}, \bibnamefont{and}
  \bibinfo{author}{\bibfnamefont{A.}~\bibnamefont{Virosztek}},
  \bibinfo{journal}{Mod. Phys. Lett. B} \textbf{\bibinfo{volume}{18}},
  \bibinfo{pages}{327} (\bibinfo{year}{2004}).

\bibitem[{\citenamefont{D\'ora and Virosztek}(2003)}]{balazs-collective}
\bibinfo{author}{\bibfnamefont{B.}~\bibnamefont{D\'ora}} \bibnamefont{and}
  \bibinfo{author}{\bibfnamefont{A.}~\bibnamefont{Virosztek}},
  \bibinfo{journal}{Europhys. Lett.} \textbf{\bibinfo{volume}{61}},
  \bibinfo{pages}{396} (\bibinfo{year}{2003}).

\bibitem[{\citenamefont{Virosztek and Maki}(1988)}]{viro2}
\bibinfo{author}{\bibfnamefont{A.}~\bibnamefont{Virosztek}} \bibnamefont{and}
  \bibinfo{author}{\bibfnamefont{K.}~\bibnamefont{Maki}},
  \bibinfo{journal}{Phys. Rev. B} \textbf{\bibinfo{volume}{37}},
  \bibinfo{pages}{2028} (\bibinfo{year}{1988}).

\bibitem[{\citenamefont{Kim et~al.}(1989)\citenamefont{Kim, Reagor, Gr\"uner,
  Maki, and Virosztek}}]{effectivemass}
\bibinfo{author}{\bibfnamefont{T.~W.} \bibnamefont{Kim}},
  \bibinfo{author}{\bibfnamefont{D.}~\bibnamefont{Reagor}},
  \bibinfo{author}{\bibfnamefont{G.}~\bibnamefont{Gr\"uner}},
  \bibinfo{author}{\bibfnamefont{K.}~\bibnamefont{Maki}}, \bibnamefont{and}
  \bibinfo{author}{\bibfnamefont{A.}~\bibnamefont{Virosztek}},
  \bibinfo{journal}{Phys. Rev. B} \textbf{\bibinfo{volume}{40}},
  \bibinfo{pages}{5372} (\bibinfo{year}{1989}).

\end{thebibliography}

\end{document}